\documentclass[prl,aps,twocolumn,showpacs]{revtex4}
\pdfoutput=1 
\usepackage{graphics}
\usepackage{graphicx}
\usepackage{amssymb,amsmath}
\usepackage{bm}

\numberwithin{equation}{section}

\newcommand{\mytitle}[1]{
 \twocolumn[\hsize\textwidth\columnwidth\hsize
 \csname@twocolumnfalse\endcsname #1 \vspace{1mm}]}

\newcommand{\beq}{\begin{equation}}
\newcommand{\eeq}{\end{equation}}
\newcommand{\bea}{\begin{eqnarray}}
\newcommand{\eea}{\end{eqnarray}}

\begin{document}

\title{Helical Majorana fermions
in
$d_{x^2-y^2}+{\it i} d_{xy}$-wave
%time-reversal broken
topological
superconductivity
of doped correlated quantum spin Hall insulators}
\author{Shih-Jye Sun$^{1}$, Chung-Hou Chung$^{2,3}$, Yung-Yeh Chang$^2$, Wei-Feng Tsai$^{4}$, and Fu-Chun Zhang$^{5,6}$
}
\affiliation{
$^{1}$Department of Applied Physics, 
National University of Kaohsiung, Kaohsiung, Taiwan, R.O.C.\\
$^{2}$Electrophysics Department, National Chiao-Tung University,
HsinChu, Taiwan, 300, R.O.C.\\
$^{3}$Physics Division, National Center for Theoretical Sciences, HsinChu, Taiwan, 300 R.O.C.\\
$^{4}$Department of Physics, National Sun Yat-Sen University, Kaohsiung, Taiwan, R.O.C.\\
$^{5}$Department of Physics, Zhejiang University, Hangzhou, China\\
$^{6}$ Collaborative Innovation Center of Advanced Microstructures, Nanjing, China
}
\date{\today}

\begin{abstract}
Large Hubbard $U$ limit of the Kane-Mele model
on a zigzag ribbon of honeycomb lattice near half-filling is studied via a renormalized
mean-field theory. The ground state exhibits time-reversal symmetry (TRS) breaking
$d_{x^2-y^2} + {\it i}d_{xy}$-wave superconductivity. At large spin-orbit coupling, the $Z_2$ phase with non-trivial spin Chern number in the pure Kane-Mele model is persistent into the TRS broken state (called ``spin-Chern phase''), 
and has two pairs of counter-propagating helical Majorana modes at the edges. As the spin-orbit coupling is reduced, the system undergoes a topological quantum phase transition from the spin-Chern to chiral superconducting states. Possible relevance of our results to adatom-doped graphene and irridate compounds is discussed.
\end{abstract}

\maketitle

%%%%%%%%%%%%%%%%%%%%%%%%%%%%%%%%%%%%%%%%%%%%%%%%%%%%%%%%%%%%%%%%%%%%%%%
{\it Introduction.}
Searching for topological states of quantum matters constitutes
one of the central
and fundamental issues in condensed matter systems.
The growing interest in topological insulators (TIs),
which support gapless
edge (or surface) states protected by time-reversal symmetry (TRS)
while the bulk remains insulating 
\cite{Kane-review,SCZhang-review}, is one prime example. 
Of particular interest are topological
superconductors which support gapless self-conjugate,
charge-neutral fermionic quasi-particle excitations \cite{alicea}.
These excitations which reflect 
non-trivial topological bulk properties are localized
at the edges, known as Majorana fermions (MFs). 
%or Majorana bound state (MBS). 

Much effort has been put in searching for
signatures of Majorana fermions in solid state materials.
One-dimensional semiconductor nano-wires with strong
SO coupling under a magnetic field
proximity to a s-wave superconductor have been proposed theoretically  
to host MF at both ends of the wire \cite{DasSarma,oppen,Flensberg}, 
and also studied experimentally \cite{delft,heiblum,Marcus,Deng,Finck}.
Similar ideas have been proposed in 2D systems where
chiral MFs exist at the edges of
spin-triplet, $p-$wave (odd-parity) superconductors 
\cite{sato,dassarma,FuKane}.

While realization of the above systems relies on TRS
breaking by the Zeeman field,
time-reversal invariant
topological superconductors (TRITOPs) \cite{XLQi,oreg,Tewari,ZhangKane} 
has recently been proposed to host
two time-reversal pairs of helical MFs at edges in
repulsively interacting SO coupled nano-wire proximity to
either a s-wave \cite{oreg,Flensberg} or a $d-$wave \cite{Law}
superconductor at each end of the wire. Proposals to realize TRITOPs 
in 2D systems include the spin-triplet
$p_x \pm {\it i} p_y$ superconductors \cite{XLQi}, the bi-layer Rashba system \cite{nakosai}, and in exciton condensates \cite{exciton}.

In this letter, we suggest a novel mechanism
for realizing helical Majorana fermions in 2D topological
superconductors with spin-singlet and TRS-breaking pairing gap-- by directly
doping correlated 2D quantum spin Hall insulators
(QSHIs or 2D TIs) \cite{Haldane,KM}.
A paradigmatic model for QSHIs is the Kane-Mele (KM) model \cite{KM,KM2},
which shows a non-trivial $Z_2$ topological (or spin Chern) number
and supports helical
topological edge states protected by TRS.
The half-filled KM model with strong electron correlations is in 
the Mott-insulating
(MI) phase \cite{hur}, while superconductivity appears upon doping.  
%We tackle the MFs in strongly correlated regime of the
%doped KM model
%via renormalized mean-field theory (RMFT) approach
%\cite{Zhang-Rice}.
 Attractive candidates to realize correlated
KM model include: graphene with enhanced SO coupling ($\sim 20meV$)
by doping with heavy adatoms, such as indium or
thallium \cite{franz}, Iridium-based honeycomb compounds
$X_2IrO_3 (X=Na$ or $Li)$
with strong SO coupling and electron correlations \cite{hur,Iridate}.

Via renormalized mean-field theory (RMFT) approach 
\cite{Zhang-Rice}, we find the spin-singlet TRS breaking
$d_{x^2-y^2}+ {\it i}d_{xy}$-wave superconductivity to appear at 
the ground state
where the chiral edge states have been shown to occur \cite{Schaffer}.
Surprisingly, for sufficiently large SO coupling,
instead of chiral edge states,
we find
zero-energy helical MFs to appear at each ribbon edge.
%protected by a distinct particle-hole symmetry.
We further show
that the system exhibits a non-trivial $Z_2$ topological invariant,  
%(or a non-trivial spin Chern number),
which supports helical MFs at edges despite the presence of TRS breaking
superconductivity. 
This seemingly un-expected feature comes as a result of
persistence of 
%$Z_2$ 
spin-Chern phase in the pure Kane-Mele model in the
 superconducting state at finite dopings. The competition between
SO coupling and chiral $d-$wave superconductivity leads to a novel
spin-Chern to chiral topological quantum phase transition.
%Due to TRS breaking superconductivity,
%the helical MFs realized in our system are distinct from
%that in TRITOPs.
%though our system at low energies
%in the limit of large  SO couplings is well
%approximated by a TRITOP.
%The relevance of our results for adatom-doped graphene and irridate compounds
%as well as their experimental signatures are discussed.\\
 %%%%%%%%%%%%%%%%%%%%%%%%%%%%%%%%%%%%%%%%%%%%%%%%%%%%%%%%%%%%%%%%%%%%%%%
\begin{figure}[t]
\begin{center}
\includegraphics[angle=0,width=8.5cm,clip]{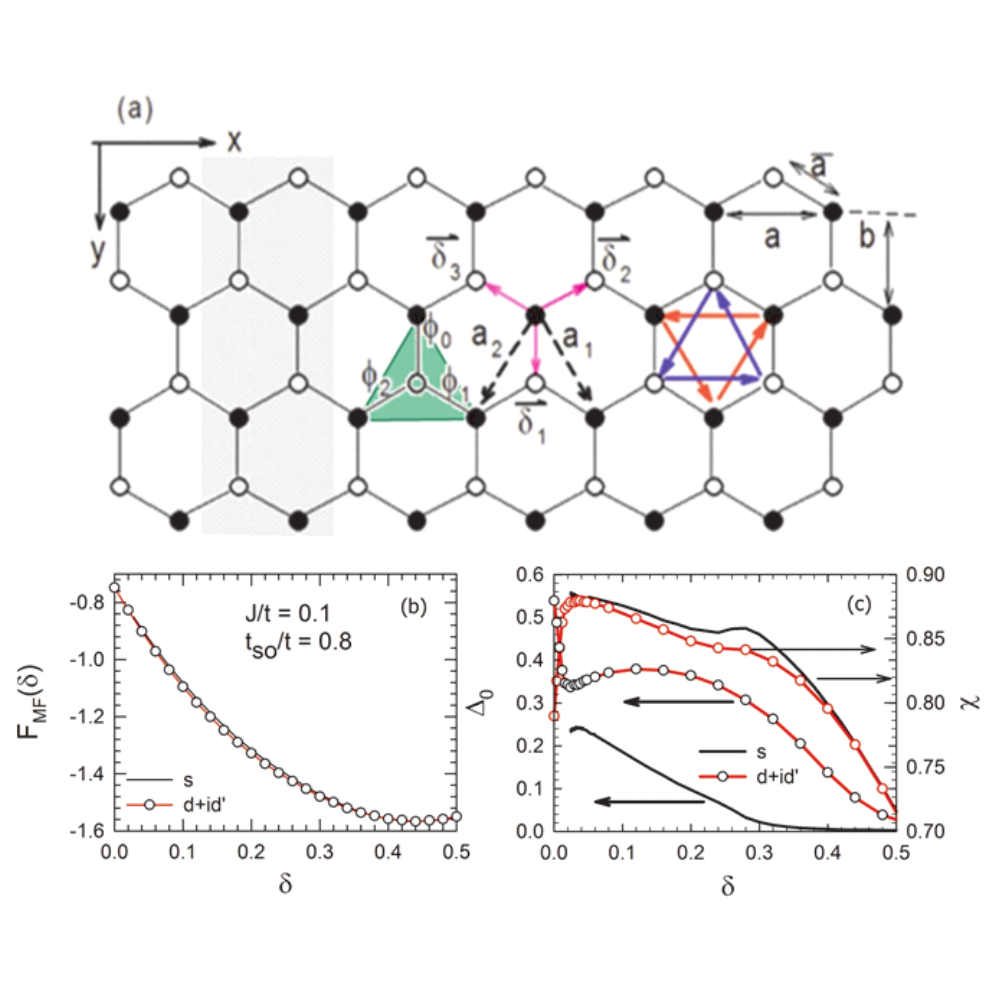}
\end{center}
\par
\vskip -0.8cm
\caption{
(Color online)
(a). Honeycomb lattice of a finite-sized zigzag ribbon of the
tight-binding Kane-Mele t-J model with the ribbon size $N=8$ being 
twice the number of zigzag chains along $x-$axis. Nearest-neighbor and
next-nearest-neighbor lattice vectors are
${\bf \vec{\delta}}_{a=1,2,3}$,
${\bf{a}}_{i=1,2}$ with an unit length of ${\bf \bar{a}},{\bf a}$, respectively.
We set ${\bf a}=1$ here. 
The gray shaded region represents for the
super-unit-cell of the zigzag ribbon, which repeats itself along $x-$axis.
The filled (open) dots stand for the sites on sub-lattice $A(B)$.
The three phases for $d+{\it i} d^\prime$ pairing gap 
are defined 
(see shaded green triangle) as: 
$\phi_{0} = 0$,$\phi_{1(2)} = -(+) 2\pi/3$. 
(b). Mean-field free energy $F_{MF}$ versus doping $\delta$
of doped Kane-Mele model on 2D periodic lattice for 
 extended $s$-wave and $d+{\it i}d^\prime$-wave superconducting pairing 
symmetries. (c). Mean-field variables $\Delta_0$ (strength of 
pairing gap) and $\chi$-field versus doping for (b). 
The parameters used
in (b) and (c) are: $J/t=0.1$, $t_{SO}/t=0.8$.
}
\label{fig3-5}
\end{figure}

{\it Model Hamiltonian.} The Hamiltonian of the Kane-Mele
t-J (KM-tJ) model is given by \cite{KM}:
\begin{eqnarray}
H &=& H_{KM} + H_J,\nonumber \\
H_{KM} &=& -t g_t\sum\limits_{\left\langle {ij} \right\rangle, \alpha }
{c_{i}^{\dagger\alpha} c_{j}^\alpha}
- \mu \sum_{i,\alpha} c^{\dagger\alpha}_i c_i^\alpha,\nonumber \\
&+&
i \frac{t_{SO}}{3} \sum\limits_{\left\langle {\left\langle {ij}
\right\rangle} \right\rangle,\alpha} {\nu _{ij} c_{i}^{\dagger\alpha} \sigma^z
c_{j}^\alpha } +  h.c.\nonumber \\
H_J &=&  J g_s \sum_{\langle i,j\rangle} (\vec{S}_i \cdot \vec{S}_j - \frac{1}{4} n_in_j)
\label{H_KM_tJ}
\end{eqnarray}
where $\alpha=\uparrow,\downarrow$ stands for the spin index, 
$\left\langle i,j \right\rangle$ and $\left\langle \left\langle i,j \right\rangle\right\rangle$ refer to the nearest-neighbor (NN) and next-nearest-neighbor (NNN) sites, respectively (see Fig.~1 (a)).
Here, $\nu_{ij}= 1$ for ${i,j}\in A$
and
$\nu_{ij}=-1$ for ${i,j}\in B$
in the SO coupling;
$\vec{S}_i$ refers to the electron spin operator on site $i$, defined as:
$\vec{S}_i=1/2 \sum_{\alpha,\beta=\uparrow,\downarrow}
c^{\dagger\alpha}_{i}\vec{\sigma}_{\alpha\beta}c_{i}^\beta$, 
$n_i=\sum_\alpha c_{i}^{\dagger\alpha} c_{i}^\alpha$ is the electron density operator,
and the anti-ferromagnetic spin-exchange couping
$J\sim \frac{4t^2}{U}$ can be derived via the
second-order perturbation from the Kane-Mele Hubbard model
in the limit of a strong on-site Coulomb repulsion $U\gg t$\cite{foot-ferro}.

The $H_J$ term has been known to favor the spin-singlet pairing.
To address superconductivity of the model,
we apply RMFT based on
Gutzwiller projected single-occupancy constraint due to
the proximity of the Mott insulating ground states,
known to well describe the ground state of $d-$wave cuprate superconductors 
%in excellent agreement with those via variational
%Monte Carlo approach
\cite{Zhang-Rice}.
As a result, the hopping $t$ term acquires a reduction factor $g_t$
at a finite doping $\delta$:
$t\rightarrow t g_t$ 
%, $t_{SO}\rightarrow t_{SO} g_t$
with $g_t = 2\delta/(1+\delta)$, while the spin-exchange
$J$ term gets enhances by a factor $g_s$: $J\rightarrow g_s J$ with
$g_s=4/(1+\delta)^2$.
The spin-exchange $J$ term within RMFT reads:
$H_{J} = H_{\chi} + H_{\Delta} + H_{const}$,
$H_{\chi} = \sum_{i,\alpha,a}
\chi_{\vec{\delta}_a} c^{\dagger\alpha}_{i}
c^{\alpha}_{i+\vec{\delta}_a} + h.c.$,
$H_{\Delta} = \sum_{i,a} \Delta_{\vec{\delta}_a}
[c^{\dagger\uparrow}_{i} c^{\dagger\downarrow}_{i+\vec{\delta}_a}
-c^{\dagger\downarrow}_{i} c^{\dagger\uparrow}_{i+\vec{\delta}_a}]
+h.c.$,
$H_{const} = N_s \sum_{a}
[\frac{|\chi_{\vec{\delta}_a}|^2}{\frac{3}{4} g_{s} J} +
\frac{|\Delta_{\vec{\delta}_a}|^2}{\frac{3}{4} g_{s} J}] - 2N_s \mu \delta$
where $a=1,2,3$, 
$N_s$ is the total number of sites,
$\chi_{\vec{\delta}_a} = -\frac{3}{4} g_{s} \sum_{\alpha}
J \langle c^{\dagger\alpha}_{i} c^{\alpha}_{i+\vec{\delta}_a}\rangle$,
$\Delta_{\vec{\delta}_a} =
-\frac{3}{4} g_{s} J \sum_{\alpha,\beta = \uparrow,\downarrow}
\epsilon_{\alpha\beta} \langle
c^{\alpha}_{i} c^{\beta}_{i+\vec{\delta}_a}\rangle$,
and $\epsilon_{\alpha\beta}={\it i}\sigma^y_{\alpha\beta}$.
Based on the $C_{6v}$ symmetry of the underlying lattice,
the pairing symmetry of
$\Delta_{\vec{\delta}_{a=1,2,3}}$ may take the following forms \cite{JPHu}:
(i) extended $s-$wave:
$\Delta_{\vec{\delta}_a} = \Delta_{\vec{\delta}_a}^s = \Delta_0$,
(ii) $d_{x^2-y^2}+{\it i}d_{xy}-$wave
(denoted also as $d+{\it i}d^\prime$):
$\Delta_{\vec{\delta}_a} = \Delta_{\vec{\delta}_a}^{d+{\it i}d^\prime} =
\Delta_0 e^{{\it i}\phi_{a-1}}$ with $\phi_0=0$, $\phi_{1(2)}=-(+)2\pi/3$ 
%with $\phi_{a-1}=2\pi (a-1) /3$ 
(see Fig. 1) \cite{hur2}.
%(iii) TRS $d \pm{\it i}d^{\prime}-$wave with
%$d+(-){\it i} d-$wave
%singlet pairing assigned to $K_+(K_{-})$ valley, respectively\cite{hur2}.
The Fourier transformed
pairing gap $\Delta_k$ for a periodic 2D lattice reads:
$\Delta_k = \sum_{a=1,2,3} \Delta_{\vec{\delta}_a}
e^{{\it i}\vec{k}\cdot \vec{\delta}_a}$.

The mean-field Hamiltonian $H_k= \psi_k^\dagger \mathcal{M}_k \psi_k$
on a periodic lattice in the basis of
$\psi_k = (c_{A,k}^\uparrow,
c_{B,k}^\uparrow,c_{A,k}^\downarrow,c_{B,k}^\downarrow,
c_{A,-k}^{\dagger\uparrow},c_{B,-k}^{\dagger\uparrow},
c_{A,-k}^{\dagger\downarrow},c_{B,-k}^{\dagger\downarrow})^T$ is given by the $8\times 8$ matrix $\mathcal{M}_k$:
\small
\begin{eqnarray}
\noindent
 \mathcal{M}_k &=&
 \left( \begin{matrix}
\hat{h}_k & \hat{\Delta}_k\\
\hat{\Delta}_k^\dagger & -\hat{h}_{-k}^\ast
\end{matrix}\right),
\hat{h}_k = \left( \begin{matrix}
{h}_k^+ & \\
 & \hat{h}_{k}^-
\end{matrix}\right),
\hat{\Delta}_k =  \left( \begin{matrix}
0 &\bar{\Delta}_k\\
-\bar{\Delta}_{k} & 0
\end{matrix}\right),\nonumber \\
 h_k^{\pm} &=& \left( \begin{matrix}
\pm\gamma_k-\mu & \epsilon_k \\
\epsilon_k^\ast & \mp\gamma_k-\mu\\
\end{matrix}\right),
\bar{\Delta}_k = \left( \begin{matrix}
0 & \Delta_k \\
\Delta_{-k} & 0\\
\end{matrix}\right)\nonumber \\
\gamma_k &=& \frac{2}{3} t_{SO} [-\sin(k_y)+
2\cos(\sqrt{3}k_x/2)\sin(k_y/2)], \nonumber \\
\epsilon_k &=& -(tg_t+\chi)\sum_{a=1,2,3}e^{{\it i}\vec{k}\cdot \vec{\delta}_{a}}.
\label{H-k-2D}
\end{eqnarray}
\normalsize
The Hamiltonian Eq.~(\ref{H-k-2D}) possesses
both the Particle-hole (PH) symmetry:
$\mathcal{C}^{-1} \mathcal{M}_{k} \mathcal{C}
= -\mathcal{M}_{-k}$,
$\mathcal{C}={\tau^x K}$
(with ${\tau^x}$ being the Pauli matrix
on particle-hole space where $K$ stands for complex conjugation 
%and $K^{-1} \mathcal{M}_k^{ij}K = (\mathcal{M}_k^{ij})^\ast$)
as well as sub-lattice symmetry                    :
$\mathcal{M}_{k}\rightarrow \mathcal{M}_{-k}$ for
$c_{A,k}\rightarrow c_{B,k}$ \cite{JPHu}.
The matrix $\hat{h}_k$,
describing the KM model, shows TRS:
$\mathcal{T}^{-1}\hat{h}_k\mathcal{T}=\hat{h}_{-k}$
where $\mathcal{T}={{\it i} \sigma^y K}$ is the time-reversal
operator taking $\left(\begin{matrix}
c_k^\uparrow,c_{k}^{\downarrow}
\end{matrix}\right)$ to
$\left(\begin{matrix}
c_{-k}^{\downarrow},-c_{-k}^{\uparrow}
\end{matrix}\right)$.
However, $\mathcal{M}_k$ breaks the TRS for $d+{\it i}d^\prime$
superconducting order parameter:
$\Delta_k^{d+{\it i}d^\prime} = \cos(\pi/3)
\Delta_{d_{x^2-y^2}}(k) + {\it i} \sin(\pi/3) \Delta_{d_{xy}}(k)$
with 
$\Delta_{d_{x^2-y^2}}(k) =
 2 \Delta_{0} (e^{-{\it i}  k_{y}/ \sqrt{3}}   
- e^{{\it i} k_{y}/2} \text{cos} (k_{x}/2 \sqrt{3}))$,  
$\Delta_{d_{xy}}(k)= -2 {\it i} \Delta_{0} e^{{\it i} k_{y}/2}  \text{sin} (k_{x}/2 \sqrt{3})$, and 
%$\Delta_{d_{x^2-y^2}}(k) =
% \Delta_{0} (   2 e^{{\it i}k_{x}/2\sqrt{3}} e^{-{\it i}k_{y}/2}- 
%e^{{\it i}k_{x}/2\sqrt{3}} e^{{\it i}k_{y}/2} - e^{-{\it i}k_{x}/\sqrt{3}})$ 
%and $\Delta_{d_{xy}}(k)=
%\Delta_0 ( e^{{\it i}k_{x}/2\sqrt{3}} e^{{\it i} k_{y}/2}-
%e^{-{\it i}k_{x}/\sqrt{3}} )$, and
$\mathcal{T}^{-1}\mathcal{M}_k^{d+{\it i}d^\prime}\mathcal{T} 
\neq \mathcal{M}_{-k}^{d+{\it i} d^\prime}$. 
%$\mathcal{T}^{-1}\Delta^{d+{\it i}d^\prime}_k\mathcal{T}=
%(\Delta^{d+{\it i}d^\prime}_{-k})^\ast \neq \Delta^{d+{\it i}d^\prime}_{k}$.
%The Bogoliubov quasi-particle
%excitation energies are found in pairs
%with two-fold degeneracy for each eigenvalue.
The mean-field free energy reads:
$F_{MF} = -\frac{2T}{N_s} \sum_{k}\ln[\cosh(\frac{E_{k}}{2T})] +
\frac{H_{const}}{N_s}$
with $E_k >0$ being positive eigenvalues and $N_s$ the number of sites.
We diagonalized the mean-field Hamiltonian $H_k$ on
a finite-sized zigzag ribbon with $N_s=N/2$ zigzag chains and
$N=56$ is set as the total number of sites along $y-$axis throughout the
paper.

{\it Results.} The mean-field variables are solved self-consistently by 
minimizing the free energy both for a periodic lattice and a finite-sized 
ribbon. On a
2D periodic lattice, the results  
as a function of doping are shown in Fig. 1(b) and
(c). Compared to the TRS
extended $s-$wave, 
%and $d_{x^2-y^2} \pm {\it i}d_{xy}-$wave
%states, 
we find $d_{x^2-y^2}+{\it i}d_{xy}-$wave
pairing is the ground state \cite{hur2}.
%$\Delta_{dx^2-y^2}(k) = \Delta_0 (e^{-{\it i}\frac{k_y}{\sqrt{3}}}
% - e^{{\it i}\frac{k_y}{2\sqrt{3}}} \cos(\frac{k_x}{2}))$ and
%$\Delta_{dxy}(k)= {\it i}\Delta_0
%e^{{\it i}\frac{k_y}{2\sqrt{3}}}\sin(\frac{k_x}{2})$
Same pairing symmetry has been reported
in superconducting phase of the doped graphene in the absence
of the spin-orbit coupling \cite{JPHu,Schaffer,baskaran,DHLee},
which was argued to support
two co-propagating chiral edge states at low energies
with a non-trivial topological winding number
$N_{TKNN}=\pm 2$
%, known as the
%class ${\bf C}$ superconductor 
\cite{blackschaffer-review}. 
The superconducting transition temperature $T_c$ is estimated
as $T_c\sim g_t \Delta_0$.

On a finite-sized zigzag ribbon and at a generic doping,
the Bogoliubov quasi-particle dispersion shows
four doubly-degenerate bulk bands 
(due to the $S_z$ symmetry of our Hamiltonian) 
grouped in two pairs (see Fig. 2(a)); it satisfies
the particle-hole symmetry 
with $2\pi$ periodicity: $E(k_x-\pi)=-E(k_x+\pi)$.
%The two pairs of Dirac-dispersed spectrum
%lines well above and below $E=0$ crossed at $k_x=\pm \pi$ 
%are originated from the helical edge states of pure 
%KM model 
%\cite{supp}. 
At low dopings, 
the normal state Fermi surfaces enclose
the Dirac points 
$K_{\pm}=(\frac{2\pi}{\sqrt{3}},\pm\frac{2\pi}{3})$ (see Fig. 2 (b)); the 
$d+{\it i}d^\prime$ pairing strength is weak near $K_-$ \cite{hur2}.
%enclosing the six Dirac
%points (see Fig. 2 (b)).
\begin{figure}[t]
\begin{center}
%\vspace{0.2cm}
%\includegraphics[angle=0,width=8.5cm,clip]{fig2all}
\includegraphics[angle=0,width=8.5cm,clip]{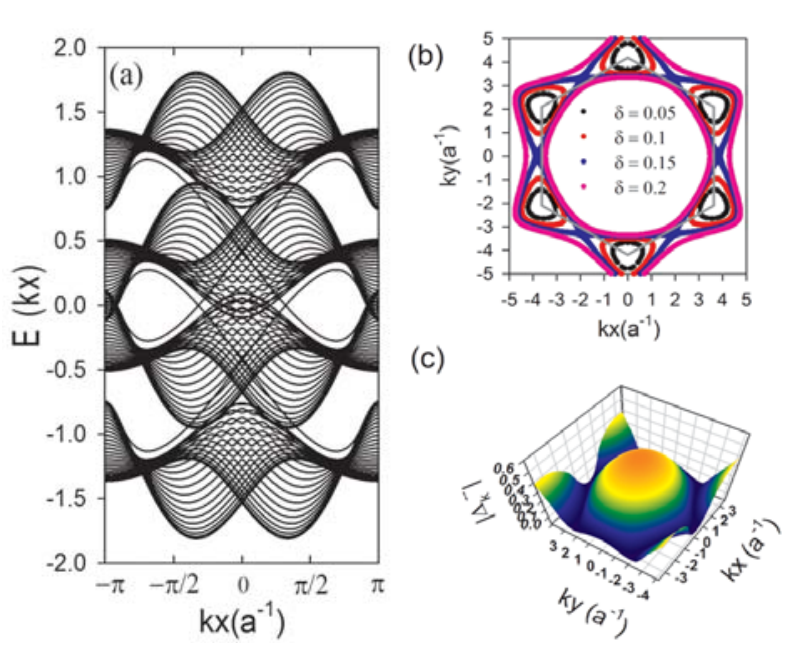}
\end{center}
\par
%\hskip -2cm
 \vskip -.5cm
\caption{
(Color online)
(a) The Bogoliubov dispersion $E(k_x)$ (in unit of $t$) 
of doped KM-tJ model on a zigzag ribbon
with $N=56$ for $J/t=0.1$,
$t_{SO}/t=0.8$, and $\delta=0.05$. (b). The spin-up Fermi surfaces
in the normal state of the Kane-Mele model on 2D periodic lattice 
for $t_{SO}/t=1$ at various dopings.
The spin-down Fermi surfaces are obtained
via $K_+\rightarrow K_-$.
(c) The 3D density plot for the
effective intra-band superconducting gap function $|\Delta^{--}_k|$ 
of the Kane-Mele model on 2D periodic lattice 
(see text and Ref.~\onlinecite{supp}).
%in Eq.~(4)\cite{supp}.
}
\label{fig7}
\end{figure}
%More interesting physics occurs near the Dirac points with enhanced
%density of states, which
%at low dopings dominate
%the low energy excitations.
%There, the pairing gap
%$\Delta^{d+{\it i}d^\prime}_{q_\pm}$
%near $K_{\pm}$ Dirac points $q_{\pm} = K_{\pm}+(\pm q_x, q_y)$ with
%$K_+= (\frac{2\pi}{3},\frac{2\pi}{3\sqrt{3}})$ and
%$K_{-}= (\frac{2\pi}{3},\frac{-2\pi}{3\sqrt{3}})$ is
%effectively a mixture of TRS $s-$wave
%(near $K_+$), dominating at high dopings
%and TRS breaking $p_x+{\it i} p_y-$wave (near $K_-$),
%dominating at low dopings and leading to chiral edge states\cite{JPHu}:
%$\Delta^{d+{\it i}d^\prime}_{q_+} \sim 3\Delta_0 e^{{\it i}4\pi/3},
%\Delta^{d+{\it i}d^\prime}_{q_-} \sim
%\frac{3}{2}\Delta_0 e^{{\it i}\pi/3} (q_y-{\it i}q_x)$.
Surprisingly, in the regime of a strong SO coupling
and weak pairing
($\Delta^{d+{\it i}d^\prime} \ll \sqrt{3}t_{SO}$),
we find the low energy excitations of our model
support helical MFs at edges instead of
chiral edge states as expected for a chiral $d-$wave superconductor.
%As shown in Fig. 3(a), 
On a finite-sized zigzag ribbon,
we find two Dirac-dispersed lines intersecting
at momenta $k^{MF}_x\sim 2\pi/3,4\pi/3$ where the Bogoliubov quasi-particle excitation energy vanishes, $E(k_x^{MF})=0$ 
(see Fig. 2(a) and Fig. 3(a))\cite{foot-nodal}.
%\cite{foot-kMF}.
%\sim \pm 2\pi (1-\frac{1}{\sqrt{3}}), 
%\pm \frac{2\pi}{\sqrt{3}}, \pm\pi$
%linked to $K_-$ ($mod ~2\pi$) point with soft pairing gap, and 
%The points at $k_x^{MF}\sim \pi/2,3\pi/2$ 
%show enhanced SO coupling and suppressed $d+{\it i}d^\prime$ pairing strength, 
%while $k_x^{MF}=\pi$ corresponds to the TR invariant $\Gamma$ point \cite{KM}. 
Near each of these gapless points, 
two pairs of two-fold degenerate states are generated via
intersecting the two Dirac lines by a constant energy
at two momentum points $k^{MF}_{x1(x2)}$, denoted as: $\Psi_{1(2),j=1,2}^{MF}$.
These four degenerate states are located at the same edge. 
However, $\Psi^{MF}_{1,j}$ and $\Psi_{2,j}^{MF}$
are counter-propagating, while 
%the degenerate states 
$\Psi_{j,1}^{MF}$ and $\Psi_{j,2}^{MF}$
are co-propagating 
(see Fig. 3(b)). 
In the basis of 
$\bar{\Psi}=[\Pi_ic_{i}^\uparrow,\Pi_ic_{i}^\downarrow,
\Pi_i c_{i}^{\dagger\uparrow},\Pi_i c_{i}^{\dagger\downarrow}]$ 
%$\bar{\Psi}=[\Pi_i(c_{A,i}^\uparrow,
%c_{B,i}^\uparrow),\Pi_i(c_{A,i}^\downarrow,c_{B,i}^\downarrow),\Pi_i
%(c_{A,i}^{\dagger\uparrow},c_{B,i}^{\dagger\uparrow}),\Pi_i
%(c_{A,i}^{\dagger\downarrow},c_{B,i}^{\dagger\downarrow})]$ 
with $i=1\cdots N$, 
%being the ribbon index along $y-$axis,
the square magnitudes 
%(upon summing over sub-lattices)
of two pairs of degenerate eigenstate wave-functions associated with the same 
$k_x^{MF}$, defined as 
$|\Psi(i)|^2 \equiv |\Psi^{R/L}_{j=1,2}(i)|^2\equiv 
(|u(i)|^2,|\tilde{u}(i)|^2,|\tilde{v}(i)|^2,|v(i)|^2)_{j}^{R/L}$, exhibit an exponential decay from both edges into the bulk, where 
%with $i=(p,p+N/2,p+N,p+3N/2)=1\cdots 2N$  
%for $E(k\sim k_x^{MF})\sim 0$ 
$R/L$ refers to the right/left moving state, 
and $u(i),\tilde{u}(i),\tilde{v}(i),v(i)$ 
are the corresponding matrix elements. 

These features are clearly different from the
co-propagating chiral edge states realized either in the chiral
$d$-wave superconductivity in doped graphene or
by proximity of a $s-$wave superconductor to a quantum anamolous Hall
insulators \cite{qixiaoliang}. Instead, the edge states we find 
fit well to the helical MFs
%(see below)
%except that they are not protected by TRS as in the case of TRITOPs.
%To show that our edge states are helical MFs protected by the
%$\tilde{\mathcal{C}}$ symmetry,
%we describe our edge states by
described by the linearly-dispersed Hamiltonian defined by
the Bogoliubov quasi-particle operators $\gamma_k^{R(L)\tau}$
as:
$H_{edge} = \sum_{\bar{k},\tau=\uparrow,\downarrow} |\bar{k}|
(\gamma_{\bar{k}}^{\dagger R\tau}\gamma_{\bar{k}}^{R\tau} -
\gamma_{\bar{k}}^{\dagger L\tau}\gamma_{\bar{k}}^{L\tau})$, $
\gamma_{k}^{R\tau} =
u_{k,\bar{i}}^{\tau} c_{k,\bar{i}}^\uparrow +
\tilde{u}_{k,\bar{i}}^{\tau} c_{k,\bar{i}}^{\downarrow}+
\tilde{v}_{k,\bar{i}}^{\tau} c_{-k,\bar{i}}^{\dagger\uparrow}+
v_{k,\bar{i}}^{\tau} c_{-k,\bar{i}}^{\dagger\downarrow}$,
 $\gamma_{k}^{L\tau} =
-v_{k,\bar{i}}^{\tau} c_{-k,\bar{i}}^\uparrow +
\tilde{v}_{k,\bar{i}}^{\tau} c_{-k,\bar{i}}^{\downarrow}
-\tilde{u}_{k,\bar{i}}^{\tau} c_{k,\bar{i}}^{\dagger\uparrow}
+ u_{k,\bar{i}}^{\tau} c_{k,\bar{i}}^{\dagger\downarrow},$
%\begin{eqnarray}
%H_{edge} &=& \sum_{\bar{k},\tau=\uparrow,\downarrow} |\bar{k}|
%(\gamma_{\bar{k}}^{\dagger R\tau}\gamma_{\bar{k}}^{R\tau} -
%\gamma_{\bar{k}}^{\dagger L\tau}\gamma_{\bar{k}}^{L\tau}), \\
%\gamma_{k}^{R\tau} &=&
%u_{k,i}^{\tau} c_{k,i}^\uparrow +
%\tilde{u}_{k,i}^{\tau} c_{k,i}^{\downarrow}+
%\tilde{v}_{k,i}^{\tau} c_{-k,i}^{\dagger\uparrow}+
%v_{k,i}^{\tau} c_{-k,i}^{\dagger\downarrow}, \nonumber\\
%\gamma_{k}^{L\tau} &=&
%-v_{k,i}^{\tau} c_{-k,i}^\uparrow +
%\tilde{v}_{k,i}^{\tau} c_{-k,i}^{\downarrow}
%-\tilde{u}_{k,i}^{\tau} c_{k,i}^{\dagger\uparrow}
%+ u_{k,i}^{\tau} c_{k,i}^{\dagger\downarrow}, \nonumber \\
%\label{H-edge} \nonumber
%\end{eqnarray}
where $\bar{k}=k-k_x^{MF}$, $k\equiv k_x$, $\gamma_{\bar{k}}^{\alpha\tau}$ with
$\alpha=L,R$
refers to the Bogoliubov quasi-particle destruction operator defined
by the coherence factors $u_{k,\bar{i}}^{\tau},\tilde{u}^{\tau}_{k,\bar{i}},\tilde{v}^{\tau}_{k,\bar{i}},v_{k,\bar{i}}^{\tau}$,
corresponding to the right-moving quasi-particle
with ``pseudo-spin''
$\tau=\uparrow(\downarrow)$ 
%corresponding to
%the two eigenvalues of $\tilde{\mathcal{C}}$,
, and the
summation over repeated site index $\bar{i}=1,\cdots N$
%along $y-$axis
is implied; similarly for $\gamma_{\bar{k}}^{L\tau}$.
The pair of the degenerate
wavefunctions $\Psi^{R(L)}_{j=1(2)}(\bar{i})$
at $k_{x1}^{MF}(k_{x2}^{MF})$
%, living on the opposite edges,
can be expressed as
$\Psi^{R(L),\uparrow(\downarrow)}(\bar{i})$,
formed by the
coherence factors:
$\Psi^{R,\uparrow}(\bar{i})=(u_{k,\bar{i}}^{\uparrow},\tilde{u}^{\uparrow}_{k,\bar{i}},\tilde{v}_{k,\bar{i}}^{\uparrow},v^{\uparrow}_{k,\bar{i}},)$,
$\Psi^{L,\downarrow}(\bar{i})=(-v_{k,\bar{i}}^{\downarrow},\tilde{v}^{\downarrow}_{k,\bar{i}},-\tilde{u}^{\downarrow}_{k,\bar{i}},u_{k,\bar{i}}^{\downarrow})$
%$\bar{i} = (i,i+N,i+2N,i+3N) = 1\cdots4N$; 
; similarly for the other doublet $\Psi^{R(L),\downarrow(\uparrow)}(\bar{i})$.
It is clear from Fig. 3 (b) that
the
edge states $(\Psi^{R(L),\uparrow(\downarrow)}
,\Psi^{L(R),\downarrow(\uparrow)})$ (as well as the Bogoliubov
operators $(\gamma_{\bar{k}}^{R\uparrow (L\downarrow)},\gamma_{\bar{k}}^{L\downarrow(R\uparrow)})$) form pairs 
%and are related
%by the sub-lattice symmetry, leading to a complete
%overlap in $|\Psi(i)|^2$ due to sub-lattice symmetry 
%upon summing over $A,B$ sub-lattice along the same 
%chain 
(see pink (blue) curve in Fig. 3(b) for
$|\Psi^{R,\uparrow}(i)|^2(|\Psi^{L,\downarrow}(i)|^2)$).
Furthermore, these Bogoliubov operators with linear dispersion
satisfy 
%the Majorana condition:
$\gamma_{-{\bf \it \tilde{k}}}(-E)=\gamma_{{\bf \it \tilde{k}}}^\dagger(E)$ with 
${\bf \it \tilde{k}}\equiv k-\pi$ via PH symmetry
(see top and bottom panels of Fig. 3(b)).
Hence, they can be regarded
as examples of helical MFs at edges 
%protected by a distinct
%$\tilde{\mathcal{C}}$ symmetry
\cite{XLQi};
the MF zero-modes occur at $k=k_x^{MF}$
(or $\bar{k}=0$)
where $\gamma_{\bar{k}=0}^{\alpha\tau}=\gamma^{\dagger\alpha\tau}_{\bar{k}=0}$.
\begin{figure}[t]
\begin{center}
\includegraphics[angle=0,width=7.5cm,clip]{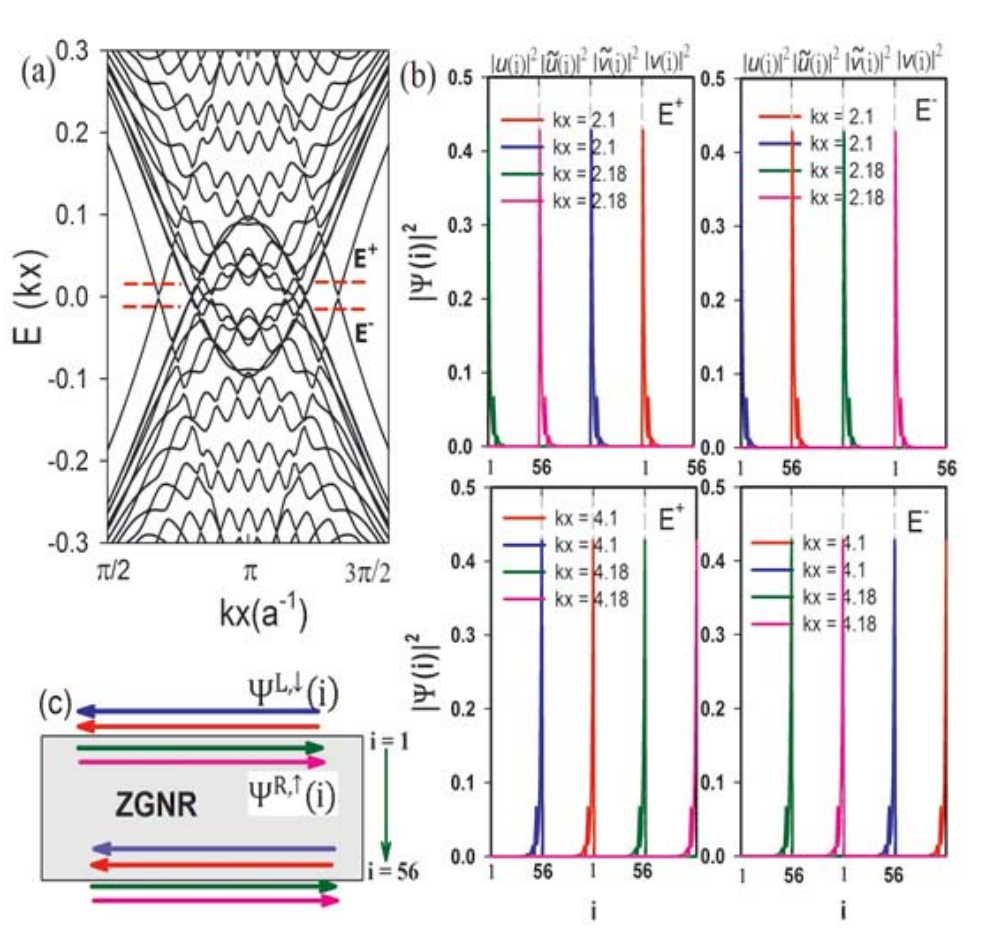}
\vskip -4.0cm
\end{center}
\par
\vskip -.50cm
\caption{
(Color online)
(a) Bogoliubov excitation spectrum of Fig. 2(a) near zero energy.
(b) The square magnitudes of two pairs of  degenerate eigen-vectors
$|\Psi^{R/L,\tau}(i)|^2\equiv 
(|u(i)|^2,|\tilde{u}(i)|^2,|\tilde{v}(i)|^2,|v(i)|^2)$ 
(see text) for a fixed 
eigenenergy $E=E^{\pm}=\pm 0.014 t$ 
with $i$ running (left to right) 
from $i=1$ (top edge) 
to $i=56$ (bottom edge),
corresponding to the helical Majorana fermions. 
%for 
%$k_{x1}= 3.10$ (top) and $k_{x2}=3.01$ (bottom) near the edges. 
Physical parameters are the same as in (a).  
(c) Schematic plot of the helical edge states in (b) for $E=E^+$ where same
color in (b) and (c) refers to the same state. Here, ZGNR refers to 
the Kane-Mele zigzag nano-ribbon.
}
\label{Fermi-surface}
\end{figure}
 These helical MFs are protected by 
an additional P-H symmetry: $\gamma_{-{\bar{k}}}(E)=\gamma_{{\bar{k}}}^\dagger(E)$ (see Fig.~3(b)). Our seemingly unexpected results have roots in the competition
between TRS SO couping and TRS breaking chiral $d-$wave superconductivity.
%The
%SO coupling leading to the $Z_2$ topological insulator in
%the original (un-doped) KM model protected by TRS. This $Z_2$
%topological insulator
We find the TRS protected
$Z_2$ QSH insulating phase of the pure un-doped Kane-Mele
model persists up to a finite
doping and a finite pairing gap in the chiral $d-$wave superconducting phase. 
%The Dirac dispersed MFs are gaped out by a Zeeman field $B_{x(y)}$,
%same as that for the pure KM model.

To gain more understanding,
we compute the $Z_2$ topological invariant (spin Chern number) 
%$N_W$ of TRITOPs
$N_W = \frac{1}{2} (C_n-C_{\bar n})$ with
$C_n=\frac{1}{2\pi {\it i}} \int_{k\in BZ}F_{12}(k)$ where the integral
is done in the first Brillouin zone (BZ),
the field strength $F_{12}(k)$ and the associated Berry's connection $A(k)$
are defined as:
$F_{12}(k)\equiv \partial_1 A_2(k) - \partial_2 A_1(k)$ and
$A_\mu = \langle n(k)|\partial_\mu|n(k)\rangle$ with $|n(k)\rangle$ being
the normalized eigenvector of the $n$th band \cite{Chern-BZ}.
$C_{\bar n}$ is defined similarly with $\bar n$ being the
%time-reversal partner
corresponding degenerate band associated with the $n$-th band 
via the transformation $S_z=\uparrow\to \downarrow$.  
%: $|\bar{n}\rangle = \mathcal{T}|n\rangle$.
In the strong SO coupling regime, $t_{SO}\gg \Delta_0$
and for a sizable range of doping around $1/2$-filling,
the 
%$Z_2$ 
spin-Chern phase prevails and we find $C_n=-C_{\bar{n}}$ or
$N_W=\pm 1$ \cite{supp},
same result as that for the $Z_2$ TRITOPs supporting helical
MFs at edges \cite{Kane-review,XLQi,foot-chern,Furusaki}. 
In this limit, 
%In the large SO coupling limit $t_{SO} \gg \Delta_0$,
our system is well approximated by the effective
spin-singlet
$p_x\pm {\it i} p_y$ superconductivity near the two Dirac points $K_\pm$.  
%resembling the case of a TRITOPs \cite{hur2,supp}.
This can be seen when re-expressing the superconducting
pairings in terms of the electron operators $\psi_{\pm,k}$ which
diagonaliz the tight-binding KM Hamiltonian 
\cite{hur2,supp}. In this basis, the intra-band pairing
$\Delta^{--}_k\psi_{-,k}^{\dagger\uparrow}\psi_{-,-k}^{\dagger\downarrow}$
 dominates at ground state (see Fig. 2 (c) and Ref.~\onlinecite{supp}).
Near $K_{\pm}$ points with $q_\pm = K_\pm + (\pm q_x+q_y)$, 
we find $\Delta^{--}_{q_\pm} \sim
\pm  \Delta_0  (q_y \pm {\it i}q_x)$, 
resembling the case of a TRITOPs. In the opposite limit for $t_{SO} \ll \Delta_0$ or sufficiently
large doping where the chiral $d$-wave pairing dominates,
however, we recover the chiral
superconductivity: $N_W=0$ and $C_n=C_{\bar{n}}=1$, equivalent to
the case of doped graphene \cite{Schaffer,JPHu}.
\begin{figure}[t]
\begin{center}
\includegraphics[angle=0,width=8.5cm,clip]{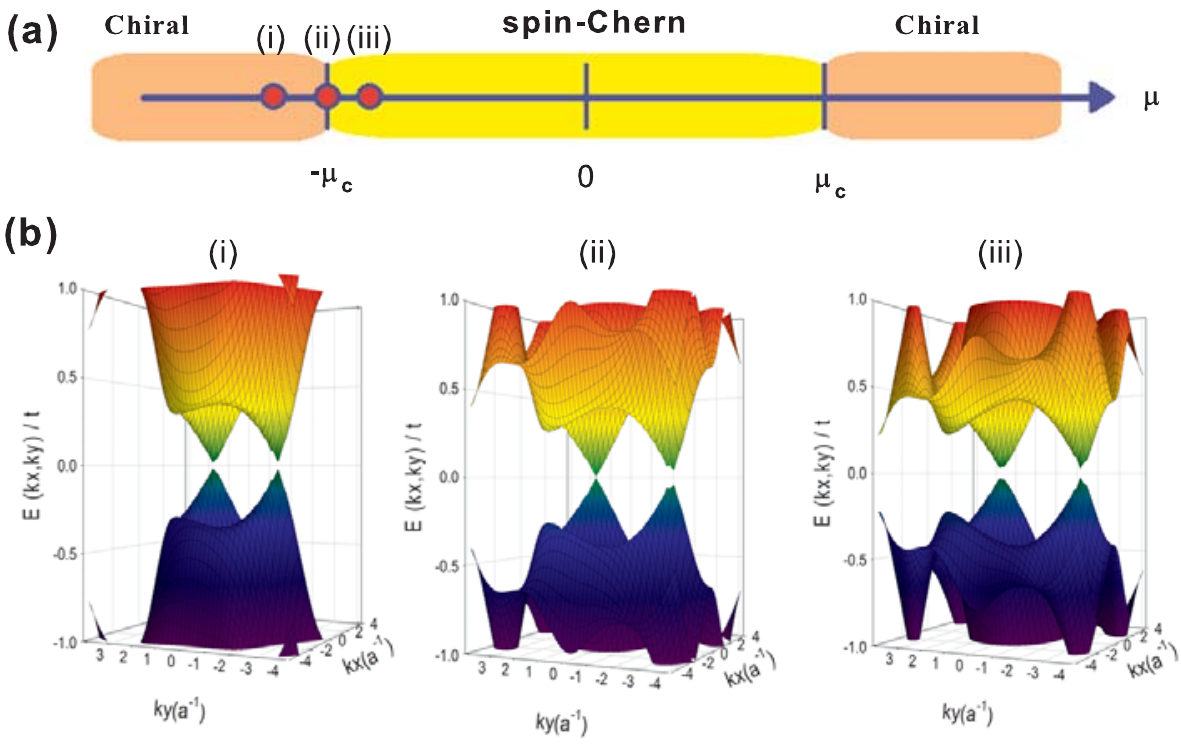}
\vskip -3.0cm
\end{center}
\par
\vskip -0.50cm
\caption{
(Color online)
(a) Topological phase diagram of doped Kane-Mele t-J model on 2D periodic
honeycomb lattice as a function of $\mu$ for fixed
$\Delta_0/t=0.3$, $t_{SO}/t=0.9$. Here, we set $\chi=0$, $g_t=g_s=1$ for
simplicity,
and $\pm\mu_c\sim \pm \sqrt{3}t_{SO}$ refers to
the critical chemical potential at the
%$Z_2$-chiral 
spin-Chern-chiral phase transition. The values of $\Delta_0$ and $\mu$ 
are tuned in a non-self-consistent way. 
(b) Energy dispersion of the two bulk bands close to zero energy.
The bulk band gap closes at the phase boundary $\mu\sim -\mu_c$ at
one Dirac point
(case (ii) with $\mu/t=-1.409,\delta=0.43$),
while it remains finite on either of the two phases (cases (i) with
$\mu/t=-1.8,\delta=0.58$ and case (iii) with $\mu/t=-1.2,\delta=0.34$).
}
\label{TQPT}
\end{figure}
A novel 
%$Z_2$-to-chiral
spin-Chern-to-chiral 
topological quantum phase transition
is found as $\Delta_0/t_{SO}$ or $\mu/t_{SO}$
is varied \cite{Mou}.
The generic phase diagram by tuning $\mu$
(in a non-self-consistent way) at a fixed $\Delta_0/t_{SO}$ is shown
in Fig.4 (a). For $\Delta_0 \ll t_{SO}$, we find
the critical values of $\mu$ being at
$\mu = \pm \mu_c \sim \pm \sqrt{3}t_{SO}$.
The bulk band gap closes only at the phase
transition, while it remains open in either phase \cite{Mou,nagaosa,FCZ}
(see Fig. 4(b)). The similar persistence of 
%$Z_2$ 
spin-Chern phase in a TRS breaking 
magnetic field has been observed experimentally in Ref.~\onlinecite{Du}. 
%Based on the symmetry classification,
The 
%$Z_2$ 
spin-Chern phase of our system belongs to class $D$
topological superconductors, distinct from the TRITOPs 
%due to the absence of
%TRS and the additional sub-lattice symmetry
\cite{Law,satonew,footnote2}. 
%which on the other hand supports helical Majorana fermion.
%This serves as an exotic and distinct
%example of topological superconductors with helical MFs distinct from the
%TRITOPs.

{\it Conclusions.}
%A promising candidate to realize our system is
%the adatom doped graphene where a enhanced
%spin-orbit coupling $t_{SO}\sim 20meV$
%compared to the pure
%graphene was predicted
%\cite{franz},
%and the on-site Coulomb interaction $U$
%is sufficiently strong ($U/t\sim 3.9$ with $t\sim 2.8eV$)
%to make it in the correlated regime
%with an estimated
%$J/t\sim 1$ to make it in the
%correlated regime\cite{blackschaffer-review}.
%To summarize,
%in contrast to the heavily studied
%chiral (helical) Majorana fermions in
%{\it spin-triplet} $p_x+{\it i} p_y$ ($p_x\pm {\it i}p_y$)
%superconductivity by either applying
%a uniform magnetic field or proximity effect,
We demonstrate here for the first time a 2D {\it spin-singlet} 
%$Z_2$
topological superconductor with non-trivial spin Chern number 
%the time-reversal-invariant topological superconductors
in doped correlated Kane-Mele model,
which supports helical counter-propagating
Majorana zero modes despite the $d+{\it i}d^\prime$
superconducting pairing gap breaks TRS.
This seemingly unexpected feature comes as a result of
persistence of 
%$Z_2$ 
spin-Chern phase of the pure Kane-Mele model in the
 superconducting state upon doping.
%the $\tilde{\mathcal{C}}$ symmetry of the mean-field Hamiltonian,
%which protects the helical MFs.
As $T\rightarrow 0$, 
distinct differential conductance spectrum for each pair of Majorana zero mode
through differential Andreev
conductance in the normal-metal/superconducting (NS)
junction is expected\cite{JPHu, Schaffer}.\\
%Further theoretical and experimental
%investigations are necessary to clarify the role of TRS
%in topological superconductors that give rise to helical MFs as
%well as the topological quantum phase transitions in our generic
%model.\\

%%%%%%%%%%%%%%%%%%%%%%%%%%%%%%%%%%%%%%%%%%%%%%%%%%%%%%%%%%%%%%%%%%%%%%%

%\subsection{ IV. Conclusions.}

\acknowledgements
We thank Y. Oreg, A. Haim, S.M. Huang, P.A. Lee, C.Y. Mou, K.T. Law, M. Sato, I.C. Fulga  
for helpful discussions.
This work is supported by the NSC grant
No.98-2918-I-009-06, No.98-2112-M-009-010-MY3, the NCTU-CTS,
the MOE-ATU program, the NCTS of Taiwan, R.O.C. (CHC), 
and NSFC grant No. 11274269 (FCZ).

%%%%%%%%%%%%%%%%%%%%%%%%%%%%%%%%%%%%%%%%%%%%%%%%%%%%%%%%%%%%%%%%%%%%%%%

%\appendix

%\section{The RG scaling equation in the weak-coupling regime}


\vspace*{-10pt}
\begin{thebibliography}{}
\vspace*{-10pt}

\bibitem{Kane-review}
M. Z. Hasan, C. L. Kane, Rev. Mod. Phys., {\bf 82}, 3045 (2010).

\bibitem{SCZhang-review}
X.L. Qi, S.C. Zhang, Rev. Mod. Phys. {\bf 83}, 1057 (2011).

\bibitem{alicea}
Jason Alicea, Rep. Prog. Phys. {\bf 75}, 076501 (2012).

\bibitem{DasSarma}
R. Lutchyn, J. Sau, and S. Das Sarma, Phys. Rev. Lett. {\bf 105}, 77001 (2010).

\bibitem{oppen}
Y. Oreg, G. Refael, and F. Von Oppen, Phys. Rev. Lett. {\bf 105},
177002 (2010).

\bibitem{Flensberg}
E. Gaidamauskas, J. Paaske, and K. Flensberg, 
Phys. Rev. Lett. {\bf 112}, 126402 (2014).

\bibitem{delft}
V. Mouril, K. Zuo, S. Frolov, S. Plissard, E. Bakkers, and L. Kouwenhoven, Science {\bf 336}, 1003 (2012).

\bibitem{heiblum}
A. Das, Y. Ronen, Y. Most, Y. Oreg, M. Heiblum, and H. Shtrikman, Nat. Phys. {\bf 8}, 887 (2012).

\bibitem{Marcus}
H. O. H. Churchill, V. Fatemi, K. Grove-Rasmussen, M. T. Deng, P. Caroff, H. Q. Xu, C. M. Marcus, Phys. Rev. B {\bf 87}, 241401(R) (2013). 

\bibitem{Deng}

M.T. Deng, C.L. Yu, G.Y. Huang, M. Larsson, P.Caro, and H.Q. Xu, Nano Lett. {\bf 12}, 6414 (2012).

\bibitem{Finck}
A.D.K. Finck, D.J. Van Harlingen, P.K. Mohseni, K. Jung, and X. Li, Phys. Rev. Lett. {\bf 110}, 126406 (2013).

\bibitem{sato}
M. Sato, S. Fujimoto, Phys. Rev. B {\bf 79} (2009).

\bibitem{dassarma}
J.D. Sau, R.M. Lutchn, S. Tewari, and S. Das Sarma, Phys. Rev. Lett. 
{\bf 104}, 040502 (2010).

\bibitem{FuKane}
L. Fu, and C.L. Kane, Phys. Rev. Lett. {\bf 100}, 096407 (2008).


\bibitem{XLQi}
X.-L. Qi, T. L. Hughes, S. R., and S.-C. Zhang,
Phys. Rev. Lett. {\bf 102}, 187001 (2009).



\bibitem{oreg}
A. Haim, A. Keselman, E. Berg and Y. Oreg, Phys. Rev. B {\bf 89}, 
220504(R) (2014).


\bibitem{Tewari}
E. Dumitrescu, and S. Tewari, Phys. Rev. B{\bf 88}, 220505(R) (2013). 

\bibitem{ZhangKane}
F. Zhang, C.L. Kane, and E.J. Mele, Phys. Rev. Lett. {\bf 111}, 056402 (2013).

\bibitem{Law}
C. L.M. Wong and K.T. Law, Phys. Rev. B {\bf 86}, 184516 (2012).


\bibitem{nakosai}
S. Nakosai, Y. Tanaka, N. Nagaosa, Phys. Rev. Lett. {\bf 108}, 147003 (2012).

\bibitem{exciton}
B. Seradjeh, Phys. Rev. B {\bf 86}, 121101(R) (2012).

\bibitem{Haldane}
F.D.M. Haldane, Phys. Rev. Lett. {\bf 61}, 2015 (1988).


\bibitem{KM}
C.L. Kane and E.J. Mele, Phys. Rev. Lett. {\bf 95}, 146802 (2005).


\bibitem{KM2}
C.L. Kane and E.J. Mele, Phys. Rev. Lett. {\bf 95}, 226801 (2005).

\bibitem{hur}
S. Rachel and K. Le Hur, Phys. Rev. B {\bf 82}, 075106 (2010).


\bibitem{franz}
C. Weeks, J. Hu, J. Alicea, M. Franz, R. Wu,
Phys. Rev. X 1, 021001 (2011).


\bibitem{Iridate}
A. Shitade, H. Katsura, J. Kunes, X.-L. Qi, S.-C. Zhang, and
N. Nagaosa, Phys. Rev. Lett. {\bf 102}, 256403 (2009);
M. Jenderka, J. Barzola-Quiquia, Z. Zhang, H. Frenzel,
M. Grundmann, and M. Lorenz, Phys. Rev. B {\bf 88}, 045111 (2013);
M. Laubach, J. Reuther, R. Thomale, S. Rachel,
arXiv:1312.2934 (un-published).



\bibitem{Zhang-Rice}
F.C. Zhang, C. Gros, T.M. Rice and H. Shiba, Supercond. Sci. Technol. {\bf 1}
36 (1988); F. C. Zhang and T. M. Rice, Phys. Rev. B {\bf 37}, 3759 (1988).


\bibitem{Schaffer}
A. M. Black-Schaffer, Phys. Rev. Lett. {\bf 109}, 197001 (2012).


\bibitem{foot-ferro}
We neglect here a small
ferromagnetic spin-exchange between NNN sites on the same sublattice 
arised from SO coupling $t_{SO}$, which may lead to spin-triplet pairing.

\bibitem{JPHu}
Y. Jiang, D.-X. Yao, E. W. Carlson, H.-D. Chen,
and J.P. Hu, Phys. Rev. B {\bf 77}, 235420 (2008).

\bibitem{hur2}
W. Wu, M.M. Scherer, C. Honerkamp, K. Le Hur, Phys. Rev. B {\bf 87}, 094521 (2013).

\bibitem{baskaran}
S. Pathak, V. B. Shenoy, and G. Baskaran, Phys. Rev. B {81}, 085431 (2010).

\bibitem{DHLee}
W.-S. Wang, Y.-Y. Xiang, Q.-H. Wang, F. Wang, F. Yang,
D.-H. Lee, Phys. Rev. B {\bf 85}, 035414 (2012).

\bibitem{blackschaffer-review}
A. M. Black-Schaffer1 and C. Honerkamp, J. Phys.: Condens. Matter {\bf 26} 
(2014) 423201.


\bibitem{foot-nodal}
Due to large $t_{SO}/\Delta_0$ ratio from our mean-field solution, 
we find bulk gap closes 
near $k_x=0,\pi$ (see Fig. 2). This comes as  
spin-orbital gap of the pure Kane-Mele ribbon gets smaller near $\Gamma$ 
point. Upon doping, the P-H symmetry of the bands is imposed, leading to 
the overlap between particle and hole bands near $k_x=0,\pi$ 
for large $t_{SO}/\Delta_0$. 
We have checked numerically that all the states near $k_x=0,\pi$ are 
indeed bulk states.  

\bibitem{supp}
See Supplementary Materials for details.

\bibitem{qixiaoliang}
X.-L. Qi, T. L. Hughes, S.-C. Zhang, Phys. Rev. B {\bf 82},
184516 (2010).

\bibitem{Chern-BZ}
T. F. Ukui, Y. H. Atsugai and H. S. Uzuki, J. Phys. Soc. Jpn.,
{\bf 74}, 1674 (2005).

\bibitem{foot-chern}
The existence of two pairs of helical edge states via mean-field analysis  
agrees perfectly with the total spin Chern number being $\pm 2$ via 
summing over $N_W$ for all four filled bands.

\bibitem{Furusaki}
A. P. Schnyder, S. Ryu, A. Furusaki, A. W. W. Ludwig,
AIP Conf. Proc. 1134, 10 (2009).



\bibitem{Mou}
The global topological phase diagram will be addressed elsewhere, see
Shin-Ming Huang, Chung-Yu Mou and Chung-Hou Chung, (in preparation).

\bibitem{nagaosa}
M. Ezawa, Y. Tanaka, N. Nagaosa, Scientific Reports, {\bf 3}, 2790 (2013).

\bibitem{FCZ}
Lunhui Hu {\it et al}, private communications.

\bibitem{Du}
L. Du, I. Knez, G. Sullivan, R.-R. Du, arXiv:1306.1925.

\bibitem{satonew}
M. Sato, Phys. Rev. B {\bf 90}, 165114 (2014).

\bibitem{footnote2}
Due to the emergent spin-singlet $p_x \pm {\it i}p_y$ 
superconductor for large spin orbit coupling 
we expect that the effect of disorder induced backscattering is reduced, 
even though it is not strictly forbidden, due to the TRS breaking 
$d+{\it i}d^\prime$ superconductivity in the background. 
As a result the localization length (along the edge) of the counter propagating 
MF helical modes will be very long for moderate disorder scattering. 
However, these helical MFs are unstable against $S_z$-preserving scatterings.


\end{thebibliography}
\end{document}

\title{Supplementary Materials for 
Helical Majorana fermions
in
$d_{x^2-y^2}+{\it i} d_{xy}$-wave
%time-reversal broken
topological
superconductivity
of doped correlated quantum spin Hall insulators}
\author{Shih-Jye Sun$^{1}$, Chung-Hou Chung$^{2,3}$, Yung-Yeh Chang$^2$, Wei-Feng Tsai$^{4}$, and Fu-Chun Zhang$^{5,6}$}
\affiliation{
$^{1}$Department of Applied Physics, 
National University of Kaohsiung, Kaohsiung, Taiwan, R.O.C.\\
$^{2}$Electrophysics Department, National Chiao-Tung University,
HsinChu, Taiwan, 300, R.O.C.\\
$^{3}$Physics Division, National Center for Theoretical Sciences, HsinChu, Taiwan, 300 R.O.C.\\
$^{4}$Department of Physics, National Sun Yat-Sen University, Kaohsiung, Taiwan, R.O.C.\\
$^{5}$Department of Physics, Zhejiang University, Hangzhou, China\\
$^{6}$ Collaborative Innovation Center of Advanced Microstructures, Nanjing, China
}
\date{\today}

\begin{abstract}
In the Supplementary Materials, we provide some of the details in the main text. 
\end{abstract}

\pacs{72.15.Qm, 7.23.-b, 03.65.Yz}
\maketitle

%%%%%%%%%%%%%%%%%%%%%%%%%%%%%%%%%%%%%%%%%%%%%%%%%%%%%%%%%%%%%%%%%%%%%%%
%{\it Introduction.} 

\section{The Hamiltonian of the Kane-Mele t-J model on a zigzag ribbon}
\begin{figure*}[t]
\centering
%\begin{center}
%\vspace{0.2cm}
\includegraphics[angle=0,width=15cm,clip]{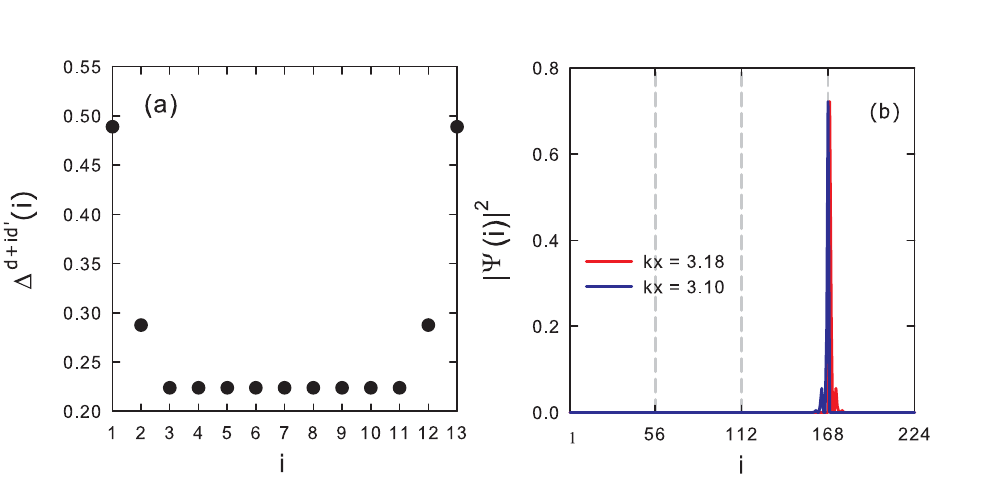}
%\includegraphics[angle=0,width=13cm,clip]{figsupp2}
%\end{center}
\par
 \vskip -0.2cm
\caption{
(Color online)
(a) Superconducting gap $\Delta^{d+{\it i}d^\prime}$ 
as a function of position in the zigzag ribbon for $N=26$ (or 
$N/2=13$ zigzag chains). (b) Square magnitude 
of wave functions $|\Psi(i)|^2$ at the edge in the superconducting states with an energy 
$E/t=-0.613$ as a function of position $i=1\cdots N$ in the zigzag 
ribbon with $N=56$. Here, we take the uniform mean-field solution for $\Delta^{d+{\it i}d'}(i) = 0.22$ for $N=56$.  
The parameters used in (a) and (b) are: $J/t=0.1$, 
$\delta = 0.15$, $t_{SO}/t=0.5$.  
}
\label{gap-site}
\end{figure*}   

Here, we provide in details the Hamiltonian of the Kane-Mele t-J model on zigzag ribbon. The Hamiltonian matrix is $4N \times 4N$ in size with $N/2$ being the
number of zigzag chains along $x$-axis. Each zigzag chain 
passes through either A
sites or B sites. The ket state of the Hamiltonian is
\begin{equation}\label{ket}
    |\Psi\rangle = (c^+_{1\uparrow} c^+_{2\uparrow}\cdots
    c^+_{N\uparrow}c^+_{1\downarrow} c^+_{2\downarrow}\cdots
    c^+_{N\downarrow}c_{1\uparrow} c_{2\uparrow}\cdots
    c_{N\uparrow}c_{1\downarrow} c_{2\downarrow}\cdots
    c_{N\downarrow})^T
\end{equation}

%\begin{figure*}[t]
%       \centering
%      % \includegraphics[scale=0.85]{HzHx005.eps}
%\includegraphics[angle=0,width=13cm,clip]{sup-fig2eps.eps}
%\vskip .50cm 
%       \caption{Bogoliubov excitation of the Kane-Mele t-J model under 
%a magnetic field $B_x/t=0.01$ along $x$-axis (a) and 
%$B_z/t=0.01$ along $z$-axis (b). The parameters used 
%here are: $J/t=0.1$, $\delta=0.15$, $t_{SO}/t=0.8$.}
%       \label{Bfield}
%\end{figure*}    

The Hamiltonian comprises three terms
\begin{equation}\label{Hal}
    H=H_0+H_{so}+H_{\Delta}.
\end{equation}
Here, $H_0$ represents the nearest-neighbor hopping terms 
in the matrix form of $(\cdots
c^+_{i\sigma}\cdots)H (\cdots c_{j\sigma}\cdots)^T$, given by:
\begin{flalign}
%\nonumber to remove numbering (before each equation)
\nonumber 
& \textrm{for} \ i = 1,3,5 \cdots N-1,  &  \\
 \nonumber 
 & \quad \quad  H_{0,\sigma\sigma}(i,i+1) = -2tcos(\frac{k}{2}), & \\
\nonumber
& \quad \quad  H_{0,\sigma\sigma}(i+1,i) = H_{0,\sigma\sigma}(i,i+1) &         \\ \\
\nonumber
 & \textrm{for} \ i = 2,4,6 \cdots N,  &       \\ 
\nonumber 
& \quad \quad H_{0,\sigma\sigma}(i,i+1) = -t,  &     \\
\nonumber 
& \quad \quad H_{0,\sigma\sigma}(i+1,i) = H_{0,\sigma\sigma}(i,i+1),  &
\end{flalign}
where $\sigma=\uparrow,\downarrow$ ($+$ or $-$), and $k=k_x$.
The matrix elements of the form $(\cdots c_{i\sigma}\cdots)H
(\cdots c^+_{j\sigma}\cdots)^T$ is related to $(\cdots
c^+_{i\sigma}\cdots)H (\cdots c_{j\sigma}\cdots)^T$ by a minus sign. 
The $H_{so}$ describes the intrinsic spin-orbit couping with the following matrix form:
\begin{flalign}
\nonumber 
& \textrm{for} \ i = 3,5,7 \cdots N-3 &\\
\nonumber
 &  \quad \quad  H_{so,\sigma \sigma}(i,i-2) = \sigma\frac{2}{3}t_{so} sin(\frac{k}{2}),  &      \\
\nonumber 
 &  \quad \quad H_{so,\sigma \sigma}(i-2,i) = H_{so,\sigma \sigma}(i,i-2),  &\\
\nonumber 
 &  \quad \quad H_{so,\sigma \sigma}(i,i+2) = \sigma\frac{2}{3}t_{so} sin(\frac{k}{2}),   &    \\
 \nonumber  &  \quad \quad H_{so,\sigma \sigma}(i+2,i)  = H_{so,\sigma \sigma}(i,i-2), & \\
\nonumber 
 &  \quad \quad H_{so,\sigma \sigma}(i,i) = -\sigma\frac{2}{3}t_{so}sin(k);    &   \\
\nonumber 
 &  \textrm{for} \  i=4,6,8 \cdots N-2 &  \\
\nonumber 
 &  \quad \quad H_{so,\sigma \sigma}(i,i-2) = -\sigma\frac{2}{3}t_{so} sin(\frac{k}{2}),  &   \\
\nonumber
  &  \quad \quad H_{so,\sigma \sigma}(i-2,i) = H_{so,\sigma \sigma}(i,i-2),    &   \\
\nonumber 
 &  \quad \quad H_{so,\sigma \sigma}(i,i+2) = -\sigma\frac{2}{3}t_{so} sin(\frac{k}{2}),   &  \\
\nonumber  
&  \quad \quad H_{so,\sigma \sigma}(i+2,i)  = H_{so,\sigma \sigma}(i,i-2),    &   \\
\nonumber  &  \quad \quad H_{so,\sigma \sigma}(i,i) = \sigma\frac{2}{3}t_{so}sin(k);   &    \\
\nonumber &  \textrm{and the other matrix elements read}  & \\
\nonumber 
 &  \quad \quad H_{so,\sigma \sigma}(1,1)=-\sigma \frac{2}{3}t_{so}sin(k),  &  \\
\nonumber  &  \quad \quad H_{so,\sigma \sigma}(2,2) = \sigma \frac{2}{3}t_{so}sin(k),   &   \\
\nonumber &  \quad \quad H_{so,\sigma \sigma}(N-1,N-1)= -\sigma\frac{2}{3}t_{so}sin(k),  &  \\
\nonumber  &  \quad \quad H_{so,\sigma \sigma}(N,N) = \sigma\frac{2}{3}t_{so}sin(k),  &  \\
\end{flalign}
where the matrix elements with the abnormal order is minus the
normal order elements. The $H_{\Delta}$ is the superconducting pairing gap 
with the matrix elements of the form 
$(\cdots c_{i,\sigma} \cdots) H (\cdots
c_{j,-\sigma})^{T} $ or $(\cdots c^+_{i,\sigma} \cdots) H (\cdots
c^+_{j,-\sigma})^{T}$, given by:
\begin{flalign}
 \nonumber 
 & \textrm{for} \ i = 1,3,5 \cdots N-1  &      \nonumber   \\
 &\quad\quad  H_{\Delta}(i,i-1+3N) = \Delta_{0} &  \nonumber  \\
 & \quad\quad H_{\Delta}(i-1,i+3N) = \Delta_{0}  &  \nonumber \\
 & \quad\quad H_{\Delta}(i,i+1+3N) = \Delta_{1}+\Delta_{2} &  \nonumber \\
&  \quad\quad H_{\Delta}(i+1,i+3N) = \Delta_{1}+\Delta_{2} & \nonumber \\
& \textrm{similarly}  &      \nonumber \\
  & \quad\quad H_{\Delta}(i+N,i-1+2N) = -\Delta_{0}   & \nonumber \\
  & \quad\quad H_{\Delta}(i-1+N,i+2N) = -\Delta_{0}  &  \nonumber \\
  & \quad\quad H_{\Delta}(i+N,i+1+2N) = -\Delta_{1}-\Delta_{2}  & \nonumber  \\
  & \quad\quad H_{\Delta}(i+1+N,i+2N) = -\Delta_{1}-\Delta_{2}  & \nonumber \\
%&  \textrm{and}  & \nonumber \\
&   \quad\quad H_{\Delta}(i+2N,i-1+N) = H_{\Delta}(i+N,i-1+2N)^{*} & \nonumber \\
 \nonumber & \quad\quad  H_{\Delta}(i-1+2N,i+N) = H_{\Delta}(i-1+N,i+2N)^{*} &  \\
 \nonumber & \quad\quad  H_{\Delta}(i+2N,i+1+N) = H_{\Delta}(i+N,i+1+2N)^{*} &  \\
 \nonumber & \quad\quad  H_{\Delta}(i+1+2N,i+N) =  H_{\Delta}(i+1+N,i+2N)^{*}&  \\
% \nonumber &  \textrm{and}  & \\
  \nonumber &\quad\quad  H_{\Delta}(i+3N,i-1) =  H_{\Delta}(i,i-1+3N)^{*} &  \\
 \nonumber  &\quad\quad  H_{\Delta}(i-1+3N,i) = H_{\Delta}(i-1,i+3N)^{*} &  \\
 \nonumber &\quad\quad  H_{\Delta}(i+3N,i+1) = H_{\Delta}(i,i+1+3N)^{*} & \\
 \nonumber &\quad\quad  H_{\Delta}(i+1+3N,i) =H_{\Delta}(i+1,i+3N)^{*}, & \\
 \nonumber & \textrm{and the other matrix elements read}  &\\
 \nonumber &\quad\quad   H_{\Delta}(1,2+3N) = \Delta_1+\Delta_2 & \\
 \nonumber &\quad\quad  H_{\Delta}(2,1+3N) = H_{\Delta}(1,2+3N) &  \\
 \nonumber & \quad\quad  H_{\Delta}(1+N,2+2N) = -H_{\Delta}(1,2+3N)& \\
 \nonumber & \quad\quad  H_{\Delta}(2+N,1+2N) =-H_{\Delta}(2,1+3N)& \\
 \nonumber &\quad\quad  H_{\Delta}(1+2N,2+N) = H_{\Delta}(1+N,2+2N)^{*}& \\
 \nonumber &\quad\quad  H_{\Delta}(2+2N,1+N) = H_{\Delta}(2+N,1+2N)^{*}& \\
\nonumber &\quad\quad  H_{\Delta}(1+3N,2) = H_{\Delta}(1,2+3N)^{*}& \\
 \nonumber &\quad\quad  H_{\Delta}(2+3N,1) = H_{\Delta}(2,1+3N)^{*}& \\
\end{flalign}
where $\Delta_0$ is the pairing order $\langle
c_{i,\sigma}c_{j,-\sigma}\rangle$, $\Delta_0$, $\Delta_1$ and
$\Delta_2$ correspond to the pairing order parameters carrying
different phases $0$,$\frac{2\pi}{3}$ and $\frac{4\pi}{3}$ in 
$d+{\it i}d^\prime$-wave
superconducting states, respectively.

%\begin{figure}[ht]
%       \centering
%       \includegraphics[scale=1.2]{magnetic.eps}
%       \caption{Bogoliubov excitation of the Kane-Mele t-J model under 
%a magnetic field $B_z$ (in unit of $t$) along $z-$axis. The parameters used 
%here are: $J/t=0.3$, $\delta=0.2$, $t_{SO}/t=1$.}
%       \label{Bfield}
%\end{figure}

\section{Additional results of the Kane-Mele t-J model on a zigzag ribbon}

\begin{figure*}[t]
       \centering
       \includegraphics[angle=0,width=10cm,clip]{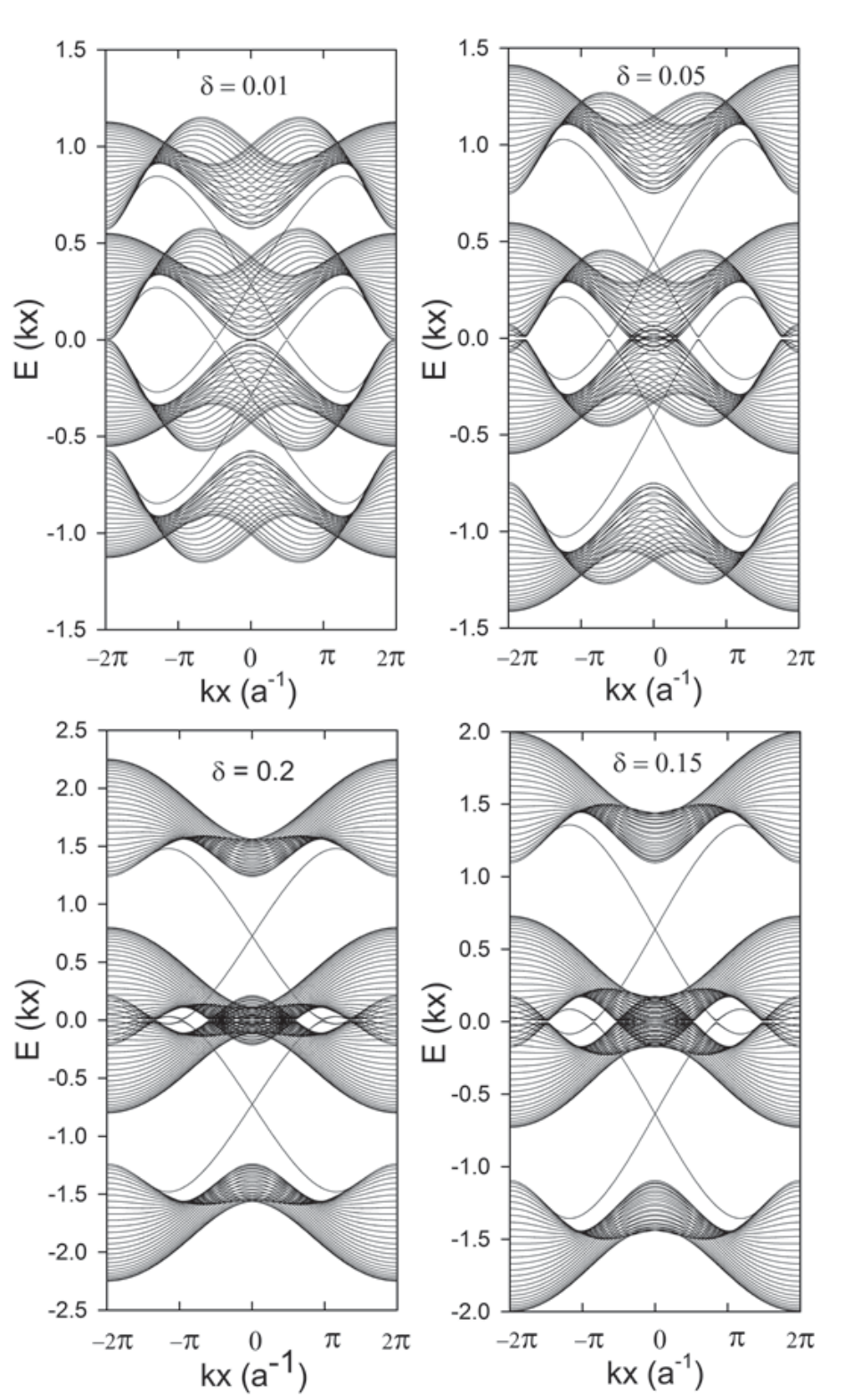}
\vskip 0.0cm 
       \caption{Bogoliubov excitation spectrum 
of the Kane-Mele t-J model for different dopings. The parameters used 
here are: $J/t=0.1$, $t_{SO}/t=0.5$.}
       \label{doping-dispersion}
\end{figure*}

Here, we provide additional results based on our numerical calculations 
on the doped Kane-Mele $t-J$ model via RMFT.

\subsection{Superconducting states at the edges}

In the Bogoliubov quasi-particle 
spectrum on the zigzag ribbon (see Fig. 2 in the main text), 
we find two linear-dispersed spectrum 
crossed at $k_x \sim \pi$ with the energy $E=\pm E_0$ ($E_0>0$). 
They are closely related to the edge states of the pure Kane-Mele model.  
The negative energy state ($E<0$) corresponds to the superconducting state 
at the edges as it supports finite values in superconducting order parameter 
$\Delta_0$ at the edges and their corresponding eigen-vectors are 
distributed mainly at the edges (see Fig.~\ref{gap-site} (b)). 

To investigate further the position dependence of the superconducting 
gap $\Delta^{d+{\it i}d^\prime}$, we generalize our mean-field calculations 
to allow for spatially varying 
superconducting gap $\Delta(i)$ (instead of the gap with 
an uniform magnitude). As shown in Fig.~\ref{gap-site} (a), 
the magnitude of gap 
is enhanced at edges and stay at a constant value in the bulk. This 
can be understood as the electronic density of states (DOS) of the un-doped 
pure KM ribbon is enhanced 
at the edges due to the presence of the helical edge states. As a result, 
the superconducting pairing strength is enhanced at edges. 
However, the inhomogeneity of superconducting gap at the edges concerns 
only with the bulk states at a sizable negative energy.    
Since we focus in this work on the Majorana fermions near zero energy, 
far away from the superconducting states at edges,  
we therefore ignore this spatial inhomogeneity and consider only 
the superconducting gap with an uniform magnitude within our RMFT approach.

%\subsection{The excitation spectrum in a magnetic field}
%
%We further study the Bogoliubov excitation spectrum of our model 
%under a magnetic field along $z$-axis $B_z$ and $x-$axis $B_x$. 
%As shown in Fig. ~\ref{Bfield} (a), 
%when the field is along $z-$axis, each of the two-fold 
%degenerate helical edge state is further splited into two states 
%without a gap opening, 
%as expected from 
%the two-fold degeneracy of the helical edge states at zero field 
%and the $S_z$ symmetry of the Kane-Mele model. 
%However, when the field is along $x$ direction, the degeneracy is lifted 
%and a gap opens at the Dirac lines, leading to the destruction 
%of the MFs (see Fig.~\ref{Bfield}(b)). The above features are qualitatively 
%the same as those of the $Z_2$ phase in the un-doped Kane-Mele model.    

\subsection{Doping evolution of the Bogoliubov excitation spectrum }

Here, we provide the doping evolution of the Bogoliubov 
excitation spectrum as shown in Fig. ~\ref{doping-dispersion}. 
The spin-Chern phase with helical Majorana modes survive 
at low dopings. For $J/t=0.1, t_{SO}/t=0.5$, 
it survives for $0<\delta<0.2$. 
Above $\delta=0.2$, the superconducting gap vanishes and  
the system is in the normal state. 
On the other hand, the chiral 
phase will prevail in the opposite regime $t_{SO}\ll \Delta_0$ 
for low dopings. However, the chiral superconductivity 
has been addressed extensively (see Refs. 29, 35 in the main text). 
We will leave this issue out of the focus of our present work.

\section{\bf  $Z_2$ topological invariant}

\begin{figure}[ht]
       \centering
       \includegraphics[scale=0.5]{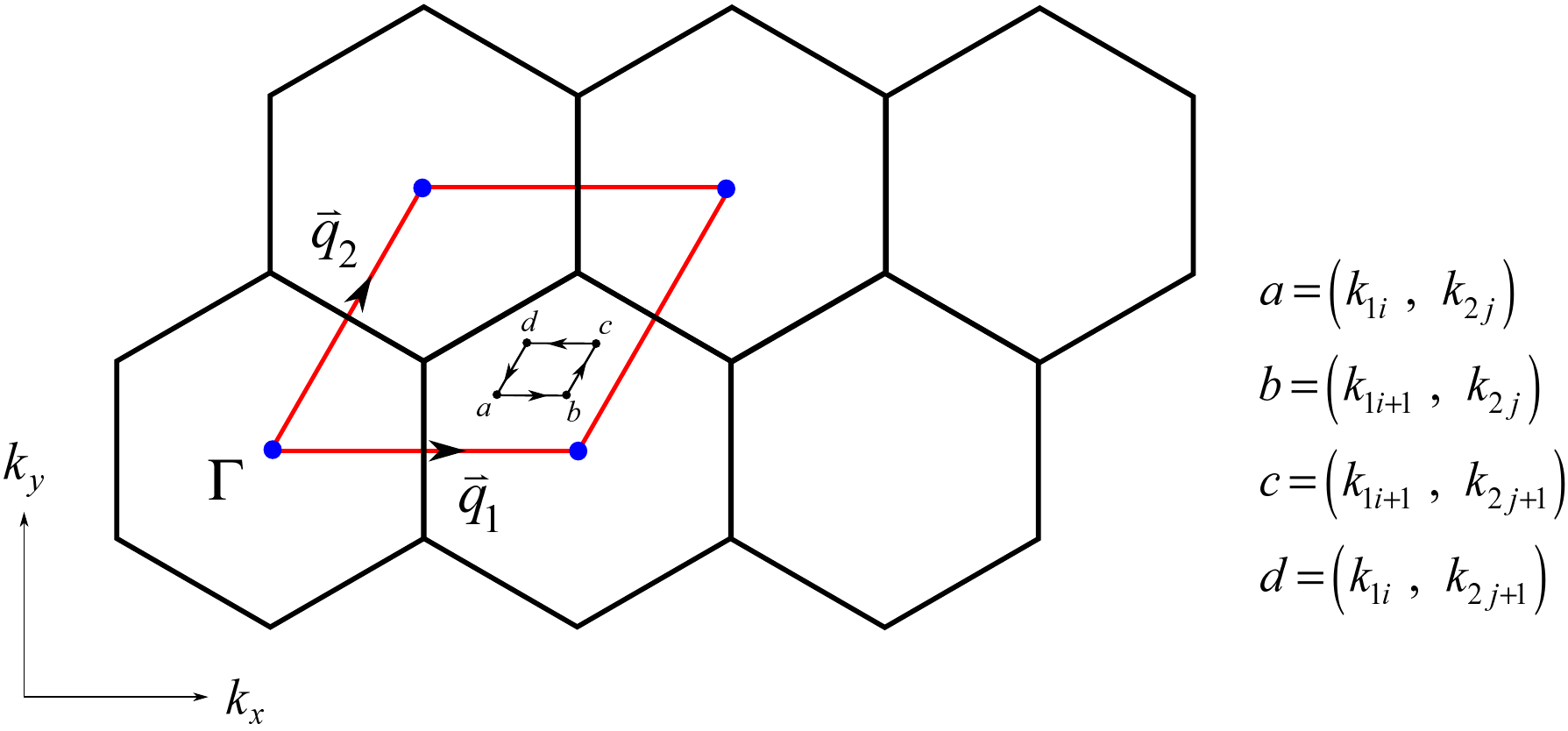}
       \caption{The parallelogram enclosed by red solid lines is the area of entire FBZ with $\Gamma = (k_{x}=0,k_{y}=0)$. The small flake of parallelogram $abcd$ is the smallest unit after discretizing the FBZ for numerical calculation.  }
       \label{fig:BZ}
\end{figure}

In this section, we provide details on obtaining the $Z_2$ topological number 
(or the spin Chern number) in the doped Kane-Mele t-J model following the approach in Ref.~\onlinecite{Chern-Japan}. 
First, the TKNN number, a topological (Chern) number, for the 
$n-$th band is defined as \cite{SCZhang-review}:
      \begin{align}
          C_{n}=\frac{1}{2 \pi i}\int_{BZ}~dk^{2}\,\, \nabla_{\bf k} \times {\bf A}({\bf k}),
          \label{eq:tknnnum}
      \end{align}
      where ${\bf A}({\bf k}) \equiv \langle n({\bf k})| {\bf \nabla}_{{\bf k}}|n({\bf k})\rangle $, $|n({\bf k})\rangle$ is the Bloch state in the $n$--th band up to a normalization coefficient, which satisfies the Schroedinger equation $H({\bf k})|n({\bf k})\rangle=E_{n}({\bf k}) |n({\bf k})\rangle $. The existence of ${\bf A}({\bf k})$ implies that there is a non-trivial magnetic field across the first Brillouin Zone (FBZ). The FBZ of the honeycomb lattice we choose for the integration Eq. (\ref{eq:tknnnum}) is illustrated in Fig. \ref{fig:BZ}, the parallelogram (area enclosed by red lines) spanned by vectors ${\bf q}_{1}$ and  ${\bf q}_{2}$ in momentum space.

      In order to numerically calculate the TKNN number covered by the entire FBZ (area enclosed by the solid red lines) in Fig. \ref{fig:BZ}, we first discretize it  into $N-1 \times N-1$ small flakes of parallelograms such as the smaller parallelogram $abcd$ (enclosed by solid black lines) in Fig. \ref{fig:BZ}. The position of a discretized lattice point (that is the vertices of the small parallelograms ) can be specified by a vector ${\bf k}_{i,j}=\left(\, k_{i1}~,~k_{j2}\right),~\text{for}~ (i,j)\in [1~,~N]$ with its components 
      \begin{align}
           & k_{i1}=\frac{1}{N-1}\left[(i-1) \times q_{1x}+(j-1) \times q_{2x} \right] ~ ; \\
            & k_{j2}=\frac{1}{N-1}\left[(i-1) \times q_{1y}+(j-1) \times q_{2y} \right],
      \end{align}
           where $q_{ix}$ and $q_{iy}$ are the components of the vectors ${\bf q}_{1}$ and  ${\bf q}_{2}$ with unit lattice spacing $|\bar{{\bf a}}|=1$ , they are given by
      \begin{align}
        & {\bf q}_{1}=\left( q_{1x}~,~q_{1y}\right)=\left(\frac{4 \pi}{3}~,~0 \right)~~;\\
        & {\bf q}_{2}=\left( q_{2x}~,~q_{2y}\right)=\left(\frac{2 \pi}{3}~,~\frac{2 \sqrt{3} \pi}{3} \right).
          \end{align}
          
     %% \begin{align}
       %%   K^{+}=\left(\frac{2 \pi}{3},~\frac{2 \pi}{3 %%\sqrt{3}} \right)~~;~~K^{-}=\left(\frac{2 \pi}{3},~-%%%\frac{2 \pi}{3 \sqrt{3}} \right)
  %%    \end{align}
      
    The phase difference $\Delta \theta_{1}({\bf k}_{ij})$ of $\Psi_{n}\left( {\bf k} \right) \equiv \langle {\bf k}|n({\bf k}) \rangle$ travelling from a lattice point ${\bf k}_{i,j}$  to ${\bf k}_{i+1,j}$ can be computed to be
    \begin{align}
       \Delta \theta_{1}({\bf k}_{ij})=&-i \, \text{ln} \, \left[\frac{\Psi_{n}^{\dagger}\left( {\bf k}_{i,j} \right)\Psi_{n}\left( {\bf k}_{i+1,j} \right)}{|\Psi_{n}^{\dagger}\left( {\bf k}_{i,j} \right)\Psi_{n}\left( {\bf k}_{i+1,j} \right)|} \right]. \\ \nonumber
        \equiv & -i \, \text{ln} \, U_{1}\left({\bf k}_{ij} \right),
    \end{align}
    where
    \begin{align}
        U_{1}\left({\bf k}_{i,j} \right) \equiv \frac{\Psi_{n}^{\dagger}\left( {\bf k}_{i,j} \right)\Psi_{n}\left( {\bf k}_{i+1,j} \right)}{|\Psi_{n}^{\dagger}\left( {\bf k}_{i,j} \right)\Psi_{n}\left( {\bf k}_{i+1,j} \right)|}
    \end{align}
    
 The subscript "$1$" of $\Delta \theta_{1}$ and $U_{1}\left({\bf k}_{ij} \right)$ represents that $\Psi_{n}$ travels along the direction of ${\bf q}_{1}$. Sum over $\Delta \theta$ of the four sides of the small parallelogram $abcd$ in Fig. \ref{fig:BZ} can one obtain the total phase shift of the Bloch wavefunction on the $n$--th band after circling around that small parallelogram. Compare with Eq. (\ref{eq:tknnnum}), we define the lattice $U(1)$ gauge field strength $F^{n}_{ij}$ by   
      \begin{align}
          F^{n}_{ij} \equiv \text{ln} \,  U_{1}\left({\bf k}_{i,j} \right) U_{2}\left({\bf k}_{i+1,j} \right) U^{-1}_{1}\left({\bf k}_{i,j+1} \right) U_{2}^{-1}\left({\bf k}_{i,j} \right)
      \end{align}
   Finally, we can calculate the TKNN number on the lattice associated the $n$-th band by

%       
       % \begin{align}
       %   F^{n}_{ij}=& \label{eq:phaseaccumu}
       %   \\ \nonumber
       %  - i \, \text{ln} &~ \left[\frac{\Psi_{n}%^{\dagger}\left( {\bf k}_{i,j} \right)\Psi_{n}\left( %{\bf k}_{i+1,j} \right)}{|\Psi_{n}^{\dagger}\left( {\bf %k}_{i,j} \right)\Psi_{n}\left( {\bf k}_{i+1,j} %\right)|}
%           \frac{\Psi_{n}^{\dagger}\left( {\bf k}_{i%+1,j} \right)\Psi_{n}\left( {\bf k}_{i+1,j+1} \right)}%{|\Psi_{n}^{\dagger}\left( {\bf k}_{i+1,j} \right)%\Psi_{n}\left( {\bf k}_{i+1,j+1} \right)|}
  %         \frac{\Psi_{n}^{\dagger}\left( {\bf k}_{i%+1,j+1} \right)\Psi_{n}\left( {\bf k}_{i,j+1} \right)}%{|\Psi_{n}^{\dagger}\left( {\bf k}_{i+1,j+1} \right)%\Psi_{n}\left( {\bf k}_{i,j+1} \right)|}
%            \frac{\Psi_{n}^{\dagger}\left( {\bf k}_{i,j+1} \right)\Psi_{n}\left( {\bf k}_{i,j} \right)}{|\Psi_{n}^{\dagger}\left( {\bf k}_{i,j+1} \right)\Psi_{n}\left( {\bf k}_{i,j} \right)|}
         %  \right].
       % \end{align}        
 %$ -i \, \text{ln} ~ \left[\frac{\Psi_{n}^{\dagger}\left( {\bf k}_{i,j} \right)\Psi_{n}\left( {\bf k}_{i+1,j} \right)}{|\Psi_{n}^{\dagger}\left( {\bf k}_{i,j} \right)\Psi_{n}\left( {\bf k}_{i+1,j} \right)|} \right]$ is nothing but the phase difference of $\Psi_{n} $ travelling from the lattice point ${\bf k}_{i,j}$  to ${\bf k}_{i+1,j}$. Therefore, Eq. (\ref{eq:phaseaccumu}) is just the total phase accumulation from travelling along the four sides of the small parallelogram. Finally, the TKNN number of the $n$--th band is defined as the summation of the phase accumulation $F^{n}_{ij}$ from the complete discretized $BZ$,           
         \begin{align}
          C_{n}=\frac{1}{2 \pi i } \sum_{i,j}F_{ij}^{n}.      
      \end{align}
   For the same band we can further define a topological number, 
so-called the spin Chern number, as:
       \begin{align}
          N_{W}=\frac{C_{n}-C_{\bar{n}}}{2},
          \label{z2numb}
      \end{align}
  where $C_{\bar{n}}$ is the TKNN number of the other degenerate eigenvector of the same band. Numerically, we obtain $C_{\bar{n}}=-C_{n}$, thus $N_{W}$ is
 either $1$ or $-1$. 
%Tab. \ref{tab:TopNum} below shows the numerical results on the calculation of the magnitude $|N_W|$ of a particular band in the doped Kane--Mele t-J model. 
We checked numerically that for the parameters used in Fig. 3 of the main text 
$|N_{W}|$ approaches to $1$. 
%$N^{2}$ is the total number of mesh points over entire FBZ ( area enclose by solid red line in Fig. \ref{fig:BZ}). \\

In the other extreme limit, $\Delta \gg t_{SO}$, we found that 
$N_{TKNN}\equiv C_n + C_{\bar{n}}=\pm 2$, 
and $N_W =0$, as expected for $d+{\it i} d^\prime$ superconductivity in doped 
graphene \cite{Mou}. 
Further studies suggest the existence of a quantum phase transition 
at a critical ratio of $\Delta /t_{SO}$ separating the 
%$Z_2$ 
spin-Chern phase 
with $N_W=\pm 1$ from the chiral (TKNN) phase with 
$N_{TKNN} =\pm 2$ (see Ref.~\onlinecite{Mou}).   
%    \\
%    \\
%\begin{table}[h]
%    \centering
%    \begin{tabular}{| c || c | c | c | }
%        \hline 
%  $N$ & 10 & 15 & 20   \\  \hline 
%    $N^{2}$ & $100 $ & $225$ & $400$  \\ \hline   
%  $|N_{W}|$ & 0.999976 & 0.999998 & 1  \\ \hline  
%  $|N_{TKNN}|$ & $8.654 \times 10^{-5}$ & $7.153 \times 10^{-5}$ & $6.629 \times 10^{-5}$  \\ \hline  
%    \end{tabular}
%    \caption{ This table shows tendency of the topological numbers 
%$N_W$ and $N_{TKNN}$ versus the increase of mesh points of $k_{x}$ and $k_{y}$ on the FBZ of a honeycomb lattice. The topological number $|N_W|$ of a particular band for the doped Kane--Mele $t-J$ model will approach to 1 while the magnitude of the TKNN number $|N_{TKNN}|$ goes to zero. The parameters used are: $t_{SO}/t=1$, $\delta=0.2$, $\chi/t=0.95$, $\mu/t=-0.88$ and $\Delta_{0}/t=0.33$.}
%    \label{tab:TopNum}
%\end{table}

\section{\bf Effective TRS superconductivity near the Dirac points}
 
In this section, we derive the low-energy effective TRS superconducting pairing
 of our model near Dirac points.  
We re-express the $d+{\it i} d^\prime$ superconducting 
order parameter in terms of the electron operators $\psi_{\pm,k}$ which 
diagonalizes the 
tight-binding KM Hamiltonian with the corresponding eigenvalue 
$E^{\pm} = \pm (\sqrt{\epsilon_k^2+\gamma_k^2}-\mu)$ (see Ref.~\onlinecite{hur2}):
\begin{equation}
 \left(\begin{array}{c}
c_{A,k}^\uparrow \\ c_{B,k}^\uparrow\\ 
\end{array}\right)
= \left( \begin{matrix}
-\alpha^-_k& -\alpha^+_k\\
\beta^-_k& \beta^+_k
\end{matrix}\right) 
 \left(\begin{array}{c}
\psi_{+,k}^\uparrow \\ \psi_{-,k}^\uparrow
\end{array}\right)
\label{trans-up}
\end{equation}
and 
\begin{equation}
 \left(\begin{array}{c}
c_{A,k}^\downarrow \\ c_{B,k}^\downarrow\\ 
\end{array}\right)
= \left( \begin{matrix}
\alpha^+_k& \alpha^-_k\\
\beta^+_k& \beta^-_k
\end{matrix}\right) 
 \left(\begin{array}{c}
\psi_{+,k}^\downarrow \\ \psi_{-,k}^\downarrow
\end{array}\right)
\label{trans-down}
\end{equation}
where $\alpha^{\pm}_k = \beta^{\pm}\times 
\frac{\epsilon_k (\gamma_k 
\pm \sqrt{\epsilon_k^2+\gamma_k^2})}{|\epsilon_k|^2}$, 
$\beta^\pm = \frac{|\epsilon_k|}{\sqrt{|\epsilon_k|^2+  
(\gamma_k \pm \sqrt{\epsilon_k^2+\gamma_k^2})^2}}$, and 
$(\alpha^{\pm}_k)^\ast = - \alpha^{\mp}_{-k}$, $\beta^{\pm} = |\alpha^{\mp}_k|$. 
The $d+{\it i} d^\prime$ 
superconducting pairing term can be re-written as:
\begin{align}
H_{\Delta} & = \sum_k \Delta_k^{d+{\it i}d^\prime} 
(c_{A,k}^{\dagger\uparrow}c_{B,k}^{\dagger\downarrow}- 
c_{A,k}^{\dagger\downarrow}c_{B,k}^{\dagger\uparrow}) + h.c.\nonumber \\
&=\sum_k [\Delta^{--}_k
\psi_{-,k}^{\dagger\uparrow}\psi_{-,-k}^{\dagger\downarrow}
+ \Delta^{++}_k
\psi_{+,k}^{\dagger\uparrow}\psi_{+,-k}^{\dagger\downarrow}\nonumber \\
&+
\Delta^{+-}_k
\psi_{+,k}^{\dagger\uparrow}\psi_{-,-k}^{\dagger\downarrow}
+
\Delta^{-+}_k
\psi_{-,-k}^{\dagger\uparrow}\psi_{+,k}^{\dagger\downarrow}] + h.c.,\nonumber \\
\Delta^{++(--)}_k &=|\alpha^{+(-)}_k||\alpha^{-(+)}_k|\nonumber \\
& \times [\Delta_k^{d+{\it i}d^\prime} e^{{\it i}\theta(\alpha^{+(-)}_k)} + 
\Delta_{-k}^{d+{\it i}d^\prime} e^{-{\it i}\theta(\alpha^{+(-)}_k)}],\nonumber \\
\Delta^{+-(-+)}_k &=\pm |\alpha^{+(-)}_k|^2e^{{\it i}\theta(\alpha^{+(-)}_k)}
[\Delta_k^{d+{\it i}d^\prime}- \Delta_{-k}^{d+{\it i}d^\prime}],\nonumber \\
\label{Delta+-}
\end{align}
where $\alpha^{\pm}_k= |\alpha^{\pm}_k| e^{{\it i}\theta(\alpha^{\pm}_k)}$ 
with $\theta(\alpha^{\pm}_k)$ being th ephase of $\alpha^{\pm}_k$. 
In the presence of a finite SO coupling and hole-doping, the 
intra-band pairing of the lower band 
$\psi_{-,k}^{\dagger\uparrow}\psi_{-,-k}^{\dagger\downarrow}$ with 
pairing gap function $\Delta^{--}_k$ dominates. 
The gap function 
$\Delta^{--}_{q_{\pm}}$ 
near Dirac points $K_{\pm}= 
(\frac{2\pi}{3\sqrt{3}},\pm\frac{2\pi}{3})$ with 
$q_{\pm} = K_{\pm}+ (\pm q_x, q_y)$ 
behaves to leading order in $|q|$ as an effective spin-singlet  
$p_x\pm {\it i} p_y$-like superconductivity: 
\begin{eqnarray}
\Delta^{--}_{q_\pm} &\sim& 
 -a_0\Delta_0 {\it i} (q_x \mp {\it i}q_y)
\label{D-minus} 
\end{eqnarray}
with $a_0=\frac{3t e^{{\it i}\pi/3}}{12\sqrt{3}t_{SO}}$. 
Hence, our system behaves as an effective TRITOPs. In deriving 
Eq.~(\ref{D-minus}), we have used the following results:
\begin{eqnarray}
\epsilon_{q_\pm} &\approx& \pm \frac{3t}{2}(q_y+ {\it i}q_x),\nonumber \\
\alpha^-_{q_-} &\approx& -\frac{\epsilon_{q_-}}{2\sqrt{3}t_{SO}},\nonumber \\
\alpha^+_{q_-} &\approx& \epsilon_{q_-} / |\epsilon_{q_-}|.
\label{approx}
\end{eqnarray}
We have also used the approximated forms of $d+{\it i} d^\prime$ pairing gap 
$\Delta^{d+{\it i}d^\prime}(q)_{\pm}$  
near the Dirac points, shown to be 
effectively a mixture of $s-$wave like (near $K_+$) 
and $p_x+{\it i} p_y-$wave like (near $K_{-}$) 
superconducting pairing \cite{JPHu}: 
\begin{eqnarray}
\Delta^{d+{\it i}d^\prime}(q_+) &\approx& 3\Delta_0 e^{{\it i}4\pi/3},\nonumber \\
\Delta^{d+{\it i}d^\prime}(q_-) &\approx& 
\frac{3}{2}\Delta_0 e^{{\it i}\pi/3} (q_y-{\it i}q_x).
\label{gap-graphene}
\end{eqnarray}

%In the presence of a large spin-orbit coupling, however, we find an effective 
%TRS superconducting pairing is generated in our case. 
%The general form for the TRS and TRB parts of the 
%$d+{\it i}d^\prime-$wave pairing are given by:
%\begin{eqnarray}
%\Delta^{--}_k &=& \Delta^{TRS}_k + \Delta^{TRB}_k,\nonumber \\
%\Delta^{TRS}_k &=& |\alpha^{+}_k||\alpha^{-}_k|
% [\Delta_k^{d+{\it i}d^\prime} e^{{\it i}\theta(\alpha^{-}_k)} + 
%\Delta_{-k}^{d-{\it i}d^\prime} e^{-{\it i}\theta(\alpha^{-}_{-k})}],\nonumber \\
%\Delta^{TRB}_k &=&\Delta^{--}_k - \Delta^{TRS}_k\nonumber \\
%\label{Delta--} 
%\end{eqnarray}
%The 3D density plots of $\Delta^{--}_k$ 
%and  $\Delta^{TRS/TRB}_k$ are 
%is shown in Fig-Delta- -. 

%%%%%%%%%%%%%%%%%%%%%%%%%%%%%%%%%%%%%%%%%%%%%%%%%%%%%%%%%%%%%%%%%%%%%%%

%\appendix 

%\section{The RG scaling equation in the weak-coupling regime}